\documentclass[draft]{agujournal2019}

\usepackage{url} %this package should fix any errors with URLs in refs.
\usepackage{lineno}
\usepackage{soul}
\usepackage{epsfig}
\usepackage{float}
\usepackage{subfig}
\usepackage{mathtools}
\usepackage{tabulary,xcolor}
\usepackage{booktabs}
\usepackage{lscape}
\usepackage{multirow}
\usepackage{amssymb}
\usepackage{changepage}
\usepackage{amsthm}
\usepackage{changepage}

\usepackage{tabularx}
\usepackage{pbox}

\drafttrue

\journalname{Water Resources Research}

\begin{document}

\title{Bathymetry Inversion using a Deep-Learning-Based Surrogate for Shallow Water Equations Solvers}

\authors{{Xiaofeng Liu}\affil{1,2},
 Yalan Song\affil{1},
 Chaopeng Shen\affil{1} 
 }

\affiliation{1}{Department of Civil and Environmental Engineering, Pennsylvania State University, University Park, Pennsylvania, USA 16802.}
\affiliation{2}{Institute of Computational and Data Sciences, Pennsylvania State University, University Park, Pennsylvania, USA 16802.}

\correspondingauthor{Xiaofeng Liu}{xzl123@psu.edu}
            
\begin{keypoints}
\item A surrogate model for shallow water equations solvers using convolutional autoencoder has high fidelity.
\item Gradient-based optimization is used to perform inversion with surrogate's automatic differentiation. 
\item Physically-based regularizations on bed elevation value and slope are necessary for usable inversion results.
\end{keypoints}

\begin{abstract}
River bathymetry is critical for many aspects of water resources management. We propose and demonstrate a bathymetry inversion method using a deep-learning-based surrogate for shallow water equations solvers. The surrogate uses the convolutional autoencoder with a shared-encoder, separate-decoder architecture. It encodes the input bathymetry and decodes to  separate outputs for flow field variables. A gradient-based optimizer is used to perform bathymetry inversion with the trained surrogate. Two physically-based constraints on both bed elevation value and slope have to be added as inversion loss regularizations to obtain usable inversion results. Using the ``L-curve'' criterion, a heuristic approach was proposed to determine the regularization parameters. Both the surrogate model and the inversion algorithm show good performance. We found the bathymetry inversion process has two distinctive stages, which resembles the sculptural process of initial broad-brush calving and final detailing. The inversion loss due to flow prediction error reaches its minimum in the first stage and remains almost constant afterward. The bed elevation value and slope regularizations play the dominant role in the second stage in selecting the most probable solution. We also found the surrogate architecture (whether with both velocity and water surface elevation or velocity only as outputs) does not show significant impact on inversion result.
\end{abstract}

%Plain Language Summary is optional for WRR
%\section*{Plain Language Summary}

%% main text
\section{Introduction}\label{sect:intro}

%What is the topic? Why is it important? What specific problem to be solved in this work/outcontribution?
%Importance of bathymetry, difficulty to obtain accurate bathyemtry, inversion is an attractive approach. 

River bathymetry is critical for many aspects of water resources management, e.g., infrastructure planning, flood prevention, navigation and transportation through waterways, river restoration, among many others. The diverse and often inspirational water features seen in rivers, from breath-taking rapids and water falls to lowland meanders, are mainly controlled by their underlying bathymetry \cite{Robert2003,Garcia2008,Julian2018}. Many measurement techniques, ranging from as simple as a rod to acoustic sounding, and more advanced technologies such as close-range photogrammetry and remote sensing, have been used to obtain bathymetric information at a point or over an area \cite{Lane2003,ChenEtAl2019,Collins2020}. However, despite the advancements in measurement techniques, there are still technological, economical, and logistical barriers in the use of bathymetry measurement and sensing to obtain high-resolution data over wide spatial and temporal ranges \cite{Fonstad2005,Casas2006,LeeEtAl2018}. 

Alternatively, indirect methods, such as bathymetry inversion from surface velocity, are attractive and have gained substantial interests  \cite{Fonstad2005,Andreadis2007,Wilson2012,LeeEtAl2018,AlmeidaWarnock2018,GhorbanidehnoEtAl2021}. This work presents one such method which uses a fast deep-learning-based surrogate model, in place of often slow physics-based models, to accelerate the inversion process. More importantly, here we employ a hyper-efficient inversion procedure based on gradient descent, automatic differentiation and backpropagation. To enable backpropagation, we make use of the fact that the neural network surrogate is already differentiable, in a similar spirit to the recently-proposed differential parameter learning (dPL) \cite{tsai2021calibration}. The gradient obtained in such automatic and fast fashion is then ingested by an inversion optimizer.

%Previous work: who has done what using which approach? Their strenght and weaknesses, which leads to our (better) approach.

Bathymetry inversion has been reported in several previous researches with very encouraging results. Like most inversion problems, the solution of bathymetry inversion is ill-posed, which is characterized by the questions of existence, non-uniqueness, and stability in response to perturbations in measurement data. Previous studies have explored the use of different approaches to obtain usable solutions and deal with the ill-posed nature of bathymetry inversion. Broadly speaking, previous works adopted either the deterministic approach with least-squares \cite{AlmeidaWarnock2018} or the stochastic approach for example the Bayesian estimation \cite{LeeEtAl2018}. Both approaches have their strengths and weaknesses. Briefly, the stochastic approach produces a probability distribution of solutions rather than a single optimal solution from the deterministic approach. It does not need the evaluation of gradient as in the deterministic approach \cite{Ghorbanidehno2020}. However, it does require reliable prior knowledge or belief about the bathymetry. Otherwise, it is prone to producing biased models. In general, the stochastic approach is computationally more expensive than the deterministic approach, especially for inversion in high-dimensional parameter space, which is the case for bathymetry inversion \cite{Aster2013}. Methodologically, these approaches have been used to solve inversion problems in many closely related fields such as hydrogeology, seismology, and fluid mechanics \cite{Laloy2017,Zhu2021,Zhou2021}.

This work adopts the deterministic approach to invert bathymetry from flow information, which is increasingly available using acoustic, radar, and image-based technologies at high resolution and low cost. For example, surface velocity can be obtained using Lagrangian drifters, large-scale particle image velocimetry (PIV), and synthetic aperture radar (SAR) \cite{Bradley2002,Muste2008,Lewis2015,Steissberg2005,Biondi2020}. 

The feasibility of bathymetry inference from surface flow information is based on the assumption that there is a strong causal relation between the two. This assumption is generally valid, especially in shallow flows where the fluid motion is dominantly horizontal and the vertical velocity is relatively small. River flow dynamics under this condition can be described reasonably well with the two-dimensional (2D) shallow water equations (SWEs), a special case of hyperbolic partial differential equations (PDEs) which approximates the full three-dimensional motion of fluid \cite{Liu2008}. In this work, a 2D SWEs numerical solver is used to generate training data and perform inversion. 

Two-dimensional SWEs and their variants are the backbone of many models for hydrological and hydraulic predictions. These models typically use traditional methods such as the finite volume method (FVM) to discretize the governing equations on a mesh. They are often called physics-based models (PBMs), e.g., SRH-2D by U.S. Bureau of Reclamation (USBR) \cite{lai2010two}, HEC-RAS 2D by U.S. Army Corp of Engineers (USACE) \cite{brunner1995hec}, FLO-2D \cite{o2011flo}, RiverFlow2D \cite{hydronia2016riverflow2d}, MIKE 21 \cite{warren1992mike}, TUFLOW \cite{huxley2016tuflow}, LISFLOOD \cite{van2010lisflood}, PAWS \cite{Shen2010},  among many others. They have been widely used in both academic researches and practice. However, for bathymetry inversion purpose, these PBMs are computationally expensive. A popular alternative is to build surrogate model which can approximate the dynamics between model input and output, at a small fraction of the computational cost of their corresponding PBMs.

Deep learning (DL), especially deep neural network (DNN), is one of the most popular ways for building surrogate models due to its capability of capturing high-dimensional nonlinearity \cite{lecun2015deep,shen2018transdisciplinary}. Its popularity is also partially due to its ability to perform automatic differentiation. The gradient of an objective function defined on a neural network, with respect to the input, is readily available and have accuracy up to machine precision. Alternatives to obtain gradient either require numerical approximation, symbolic differentiation, or analytical derivation. For example, \citeA{AlmeidaWarnock2018} used the variational inversion method where the gradient is calculated with both forward and adjoint solutions. Although efficient, adjoint-based approach involves substantial mathematical derivations. It is also intrusive to  PBMs source code because the adjoint problem needs to solve a new set of equations. Thus, the use of automatic differentiation to obtain gradient is appealing and has been swiftly adopted by many researchers for inversion problems \cite{RenEtAl2020,Xu2020,Zhu2021}. Recently, \citeA{tsai2021calibration} proposed a novel differentiable parameter learning framework to integrate big-data DL and differentiable PBMs for parameter estimation. Here, the differentiable PBMs refer to models implemented directly using the machine learning platforms such as Tensorflow or PyTorch, or their differentiable surrogates.

%Short review of commonalityies in inversion problem, specialties in bathymetry inversion, such as existance, non-uniqueness, and stability.

%2D SWEs solvers/models: physics-based models, including SRH-2D. 

%surrogate model to SWEs solvers: who has done what? what is new in this work.

%Based on above discussion, describe some specific problems or questions to be solved/answered in this work. 

%One paragraph for the structure of this paper: methods, results, discussion, etc. 

In this work, the USBR SRH-2D solver was selected as the PBM and its surrogate was built as an autoencoder using convolutional neural network (CNN). Specifically, a shared-encoder, separate-decoder architecture is used where the input image of bathymetry is encoded and then decoded to three outputs, namely, the two flow velocity components and the water surface elevation ($WSE$). Similar structure has been used in \citeA{guo2016convolutional} for incompressible flows around objects. The training data was generated by running sufficient number of SRH-2D simulations, whose input bathymetry data was randomly generated. The simple setup in this work is sufficient to prove the concept and show the feasibility of the proposed methodology. For practical use, the training data generation should take into considerations of the prior on real bathymetry.  

This work is different from previous studies in the following aspects: (1) Our inversion is built upon a CNN-based surrogate and the differentiable parameter learning idea. Previous bathymetry inversion works, which showed very encouraging results, used different approaches such as variation inversion in \citeA{AlmeidaWarnock2018}, principal component geostatistical approach (PCGA) in \citeA{LeeEtAl2018}, and fully connected DNN in \citeA{GhorbanidehnoEtAl2021}. In fact, the approach in \citeA{GhorbanidehnoEtAl2021} directly maps surface velocity to bathymetry, i.e., there is no need for inversion. They also subdivided the domain into small segments and bathymetry inference was performed for each segment sequentially. In our work, the bathymetry inversion for the whole domain is performed in a ``one-shot'' fashion. The PCGA approach in \citeA{LeeEtAl2018} requires the transformation of the irregular domain into a rectangular box. Although our example is also on a simple rectangle domain, the input and output images used by our surrogate model can embed a river of any shape. (2) We use the gradient generated from automatic differentiation. As will be shown, we examined in detail the inversion process from the initial guess to the final converged bathymetry, which resulted in valuable insights on two distinctive stages during the inversion process. (3) To overcome the ill-posedness of the inversion problem, we found that there are two necessary physical constraints on both bed elevation value and slope, which have to be embedded in the inversion loss function as two separate regularizations. A heuristic approach is proposed to solve the problem of regularization hyperparameter determination. (4) We also investigated the effects of autoencoder architecture for surrogate model on bathymetry inversion.

The rest of the paper is organized as follows. The methodology, including the surrogate model architecture, inversion algorithm, and training data generation, is introduced first. Then the results on both the performance of the surrogate model and the inversion algorithm are presented with discussion. This paper is concluded with a summary of findings and future work. 

\section{Methodology}\label{sect:method}
This section describes the deep-learning-based surrogate model and the inversion algorithm, followed by the introduction to the PBM solver and training data generation process. 

\subsection{Deep-learning-based surrogate model architecture}\label{sect:surrogate_architecture}
\begin{figure}[htp]
\centering
    \includegraphics[width=1.0\textwidth]{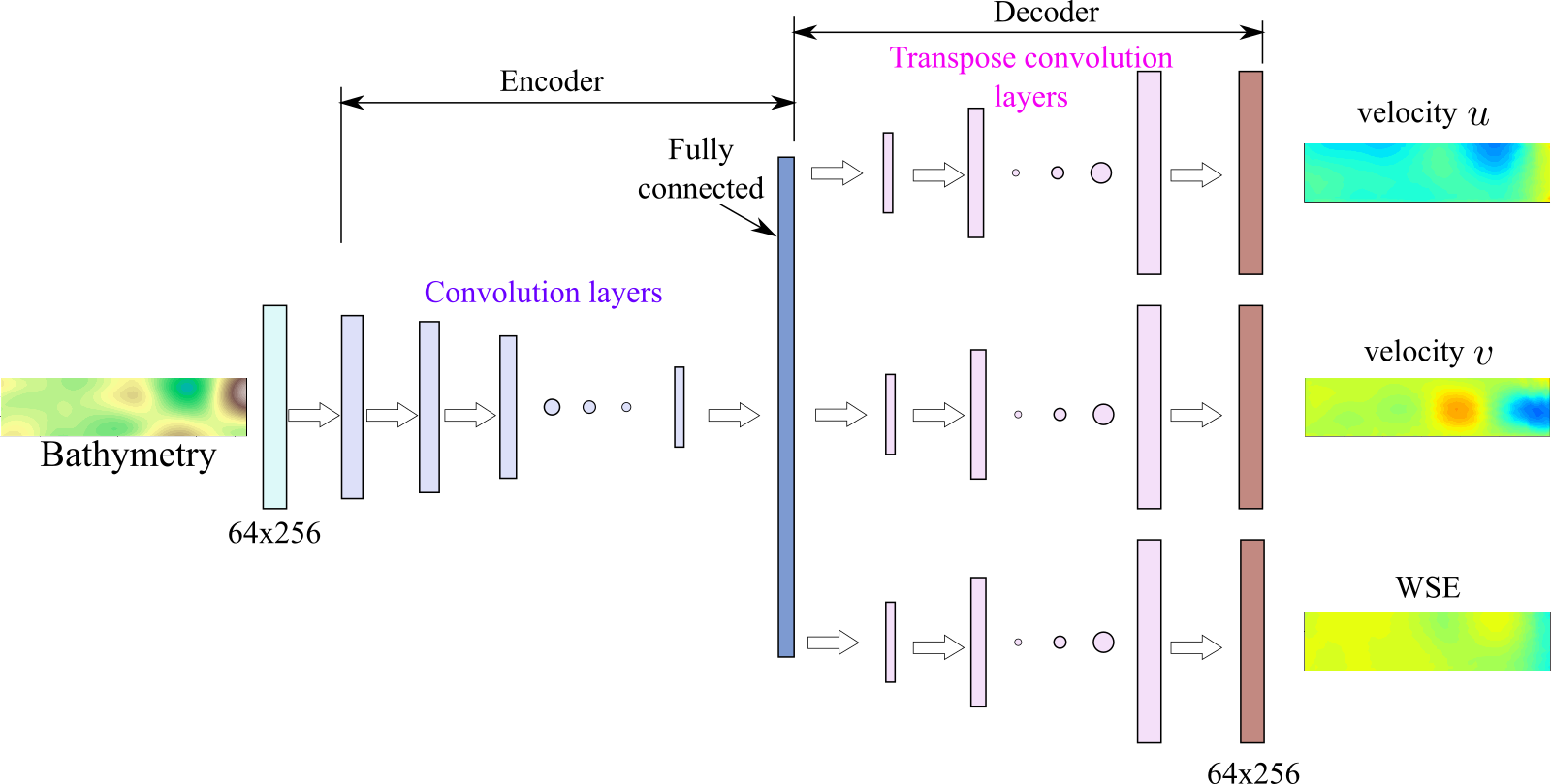}
    \caption{Scheme diagram of the CNN-based surrogate model architecture.}\label{Fig:CNN_structure}
\end{figure}

The convolutional autoencoder is used to construct the surrogate model. It consists of two parts: encoder and decoder. Figure~\ref{Fig:CNN_structure} shows the architecture of the CNN-based surrogate model developed in this work. Similar structure has been used in \citeA{guo2016convolutional} for steady-state, laminar Naiver-Stokes equations, and \citeA{Forghani2021} and \citeA{song2021surrogate} for 2D SWEs. The main differences between our surrogate model and that in \citeA{Forghani2021} (named AE in their work) are in the detailed architecture. In addition, we used multiple output branches for different flow field variables (velocity components and water surface elevation) to investigate the effects of surrogate architecture on inversion. 

The encoder takes the bathymetry image, $z_b$, as the input and gradually extracts bathymetric features through several stacked convolution layers and a fully connected layer at the end. The output values of the fully connected layer is the encoded feature vector $\mathbf{h}$, or simply code, in latent space to compactly represent the input data. The encoder can be denoted as
\begin{equation}
   \mathbf{h}(z_b) = ConvNN(z_b)
\end{equation} 
where $ConvNN(\cdot)$ is the convolutional encoding operation which maps $z_b$ image to $\mathbf{h}$ vector.

The decoder is the reverse operation of the encoder. In the original design of autoencoder, the output of the decoder should approximately reproduce the original input. However, in the framework of surrogate model for SWEs solvers, the output of the decoder is the flow field corresponding to the input bathymetry. Indeed, the decoder has three outputs, i.e, $u$, $v$, and $WSE$. Here $u$ and $v$ are velocity components. A CNN model constructed in this manner has a shared-encoder and separated-decoder architecture \cite{guo2016convolutional}. The decoder can be denoted as
\begin{equation}
  \mathbf{r}(z_b) = DeconvNN(\mathbf{h}(z_b))
\end{equation}
where $DeconvNN(\cdot)$ is the transpose convolution operation (often incorrectly noted as deconvolution and thus the name in the notation), which reconstructs the flow field from the encoded feature vector $\mathbf{h}$. 

To train the surrogate model, a proper loss function needs to be defined. The goal is to train the surrogate model to make accurate predictions on the flow field. Here, the loss function is constructed to minimize the mean squared error (MSE) of all components of the predicted flow field, i.e., $\hat{u}$, $\hat{v}$, and $\widehat{WSE}$. The hat denotes the predicted variable. It will also be used to denote the inverted variable in this work. The loss function based on the error in flow prediction can then be written as
\begin{equation}\label{eqn:L_prediction}
L_{prediction}= \frac{1}{N_{batch}}  \frac{1}{M_{sample}} \sum_{n=1}^{N_{batch}}\sum_{m=1}^{M_{sample}} \left[ (\hat{{u}}-{u})^2+ (\hat{{v}}-\overline{v})^2 + (\widehat{WSE}-WSE)^2 \right]
\end{equation} 
where $u$, $v$, and $WSE$ are the flow data (ground truth) from SRH-2D simulations. $N_{batch}$ is the number of batches and $M_{sample}$ is sample size in each batch. $n$ and $m$ are the index for batch and sample, respectively. A stochastic gradient descent optimizer, ADAM, was used to update the weights and biases of the neural network \cite{kingma2014adam}. To alleviate the vanishing gradient problem, the Rectified Linear Unit(ReLU) activation function was used \cite{sekar2019fast}. For the final output layer of the decoder, the linear activation function is used to accommodate negative flow variable values. 

The training of the surrogate model also involved hyperparameter tuning. The important hyperparameters for encoder and decoder include the number of convolution/transpose convolution layers, and within each layer the number of feature maps, filter size (width and height), strides, and padding options. The length of the encoded feature vector is also a key parameter, which determines the accuracy of latent space representation for an input bathymetry image. These hyperparameters need to be tuned to achieve the best results. Detailed explanations for each of these hyperparameters in autoencoders and their effects are omitted in this paper due to length limit. More details can be found in many deep learning textbooks, e.g., \citeA{GoodfellowEtAl2016}.

\subsection{Inversion algorithm}\label{sect:inversion_process_method}
The trained surrogate model was used for the inversion, i.e., to find the bathymetry given a flow field. For convenience, the surrogate model can be written as
\begin{equation}\label{eqn:surrogate}
   G(\mathbf{z_b}) = \mathbf{d} + \mathbf{\epsilon}
\end{equation}
where $G$ represents the forward model or physical simulator, $\mathbf{d}$ = ($u$, $v$, $WSE$) is the flow field data, and $\mathbf{\epsilon}$ is error due to numerical approximation and/or measurement inaccuracy. The inversion problem is then to find the inverse operator $G^{-1}$, which is often very challenging if the problem is ill-posed, non-linear, and high-dimensional. A key strategy to stabilize the inversion is to impose additional constraints, generally referred to as regularization. Among many, Tikhonov regularization is commonly used, which adds the additional constraint on the inverted variable. Many previous works have used the zeroth-order Tikhonov regularization where the additional constraint is to minimize the norm of $\mathbf{z_b}$. However, we will show that the zeroth-order Tikhonov regularization itself is not enough for bathymetry inversion. In this work, we propose to use two new physically-based regularizations, one for the bed elevation value (a modified zeroth-order Tikhonov regularization indeed) and one for the bed slope (a first-order Tikhonov regularization). The total inversion loss function has the form of
\begin{equation}\label{eqn:inversion_L_total}
  L_{total}(z_b) = L_{prediction} + \alpha_{value} L_{value} + \alpha_{slope} L_{slope}
\end{equation}
where $L_{prediction}$ is the flow prediction error defined in Eq.~\ref{eqn:L_prediction}, $L_{value}$ is the loss due to the inverted $\hat{z}_b$ values going beyond a prescribed range, and similarly $L_{slope}$ is the loss due to the slope of inverted bed exceeding a upper limit. For a specific river, the maximum bed slope is mostly determined by its sediment characteristics.  $\alpha_{value}$ and $\alpha_{slope}$ are the regularization factors for their corresponding losses. Like other hyperparameters in the model, these two parameters are problem specific and need to be tuned. For inverse problems, these parameters can be  determined with methods such as the L-curve criteria. Their determination and effects will be shown in the results section. 

\begin{figure}[htp]
\centering
    \includegraphics[width=0.6\textwidth]{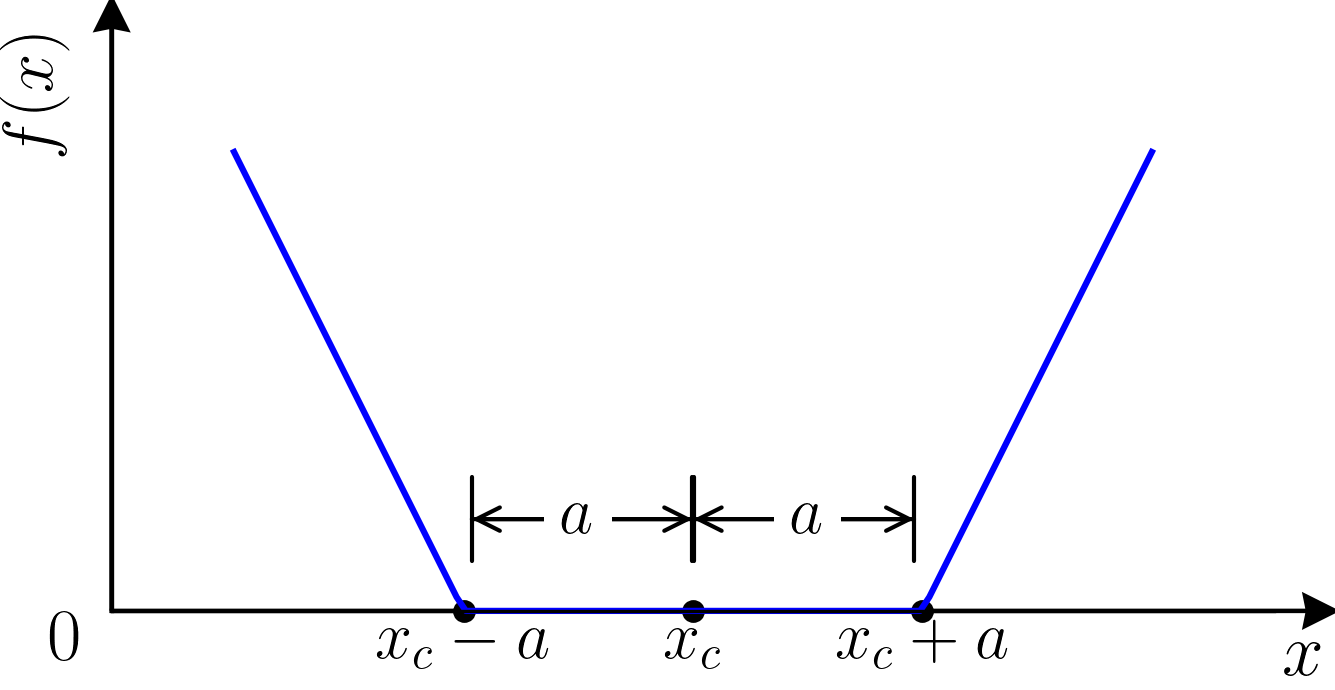}
    \caption{Double-bounded ReLU function for the calculation of inversion losses due to value and slope.}\label{Fig:double_bounded_relu}
\end{figure}

Essentially, the value loss $L_{value}$ and the slope loss $L_{slope}$ are used to discourage the inversion algorithm from searching solutions outside the prescribed bounds for bed elevation and slope. These kind of losses are commonly used in inversion problems for distributed parameters to find the most probable solution that reasonably matches the observation data. The slope loss here penalizes the total variation of bed elevation and is also referred to as the maximization of entropy. From the physics point of view, the slope of a sediment bed is bounded by the angle of repose, exceeding of which will cause sand slide \cite{Song2020}. To embed these prior information or constraints in the inversion process, the value and slope losses are calculated with a double-bounded ReLU (dbReLU) function \cite{RenEtAl2020}, which has the form of 
\begin{equation}
  f_{dbReLU}(x) = \text{ReLU}(|x-x_c| - a) = \max (0, |x-x_c| - a)
\end{equation} 
where $x$ is the input to the function, $x_c$ is the center of $x$, and $a$ is a constant which controls the range of $x$ where the loss function is zero. A schematic diagram of the function is shown in Fig.~\ref{Fig:double_bounded_relu}. It is clear that with this function as loss function, the loss will be zero within the range of $[x_c -a, x_c+a]$, i.e., the zero loss will be bounded by the two ends of the range and hence the name. Away from the two end points of this range, the loss starts to increase linearly. With the above definition, the losses due to value and slope can be written as
\begin{equation}
  L_{value} = \sum_{i=1}^{N_{points}} f_{dbReLU} (\hat{z}_b)
\end{equation}

\begin{equation}
  L_{slope} = \sum_{i=1}^{N_{points}} f_{dbReLU} (\hat{S}_x) + \sum_{i=1}^{N_{points}} f_{dbReLU} (\hat{S}_y)
\end{equation}
where $N_{points}$ is the number of points in the input bathymetry array, $\hat{S}_x$ and $\hat{S}_y$ are the slopes of inverted bed in the $x$ and $y$ directions, respectively. 

The method for bathymetry inversion proposed in this work involves two main steps. The first is to train the CNN-based surrogate model and the second is to perform the inversion. For the first step, the neural network is trained on a given set of data, which is made of the input/output pairs of the SWEs simulator. The training data generation will be described in the next section. The first step will result in a surrogate model $G$. In the second step, for any arbitrary input $z_b$, the surrogate model produces an output ($\hat{u}$, $\hat{v}$, $\widehat{WSE}$) and the inversion loss defined in Eq.~\ref{eqn:inversion_L_total} can be calculated. In the framework of neural network and due to its capability to perform automatic differentiation, the total inversion loss $L_{total}$ is differentiable with respect to the input $z_b$. The gradient $\partial L_{total}/\partial z_b$ can then be used in a simple iterative scheme to invert the bathymetry, which can be written as
\begin{equation}\label{eqn:zb_inversion}
   \hat{z}_b^{i+1} = \hat{z}_b^{i} - \alpha_{inversion} \left. \frac{\partial L_{total} \left(z_b \right)}{\partial z_b} \right\rvert_{z_b=\hat{z}_b^i}
\end{equation}
where $i$ is the iteration number and $\alpha_{inversion}$ is the inversion step size (learning rate). More advanced iterative schemes, such as the Gauss-Newton and Levenberg-Marquardt methods, can be used in future work to further take advantage of the gradient information. Equation~\ref{eqn:zb_inversion} is essentially a gradient descend optimization to minimize the total inversion loss $L_{total}$. It is noted that during the inversion process, the parameters of the surrogate neural network are frozen (not trainable). The only trainable parameters are the bed elevations $\hat{z}_b$. To start the iterations, an initial guess on the bathymetry, $\hat{z}_b^{0}$, is needed. The initial guess can be samples drawn from a prior distribution and each will result in an inverted bathymetry. The ensemble can be used to assess the stability of the inversion and calculate a mean inverted bathymetry.  

\subsection{Data generation and preprocessing}

\subsubsection{SWEs solver}
The training data for the surrogate model was generated with the 2D SWEs solver SRH-2D, which is a popular 2D hydraulics model \cite{lai2008srh}. SRH-2D is developed by the U.S. Bureau of Reclamation for many of the nation's large water resources planning and design projects. It is also adopted by the U.S. Federal Highway Administration for the design and protection of transportation infrastructure against flooding. The method developed in this work is generic and can be used with training data produced with other SWEs solvers.

The SWEs can be derived by depth-averaging the 3D Navier-Stokes equations. They have the following general form \cite{lai2010two}:
\begin{equation}\label{eqn:cty}
\frac{\partial h}{\partial t} + \frac{\partial h u}{\partial x} + \frac{\partial h v}{\partial y} =0
\end{equation}
\begin{equation}\label{eqn:momentum_x}
\frac{\partial h u}{\partial t} + \frac{\partial h u u}{\partial x} + \frac{\partial h u v}{\partial y} =  \frac{\partial h {T_{xx}}}{\partial x}+\frac{\partial h {T_{xy}}}{\partial y}-gh\frac{\partial z_s}{\partial x}-\frac{\tau_{bx}}{\rho}
\end{equation}
\begin{equation}\label{eqn:momentum_y}
\frac{\partial h v}{\partial t} + \frac{\partial h uv}{\partial x} + \frac{\partial h v v}{\partial y} =  \frac{\partial h{T_{xy}}}{\partial x}+\frac{\partial h {T_{yy}}}{\partial y}-gh\frac{\partial z_s}{\partial y}-\frac{\tau_{by}}{\rho}
\end{equation}
where $u$ and $v$ are the depth-averaged flow velocities in $x$ and $y$ directions, respectively; ${h} $ is the water depth; $z_s = h+z_b$ is $WSE$; $z_b$ is the bed elevation;  ${T_{xx}}$,${T_{xy}}$, and ${T_{xy}}$ are the depth-averaged turbulence stresses; $\tau_{bx}$ and $\tau_{by}$ are the bed shear stresses in $x$ and $y$ directions, respectively; $g$ is the gravitational acceleration; $\rho$ is the water density. More details about the turbulence model and the physical means of all terms can be found in \citeA{rodi1993turbulence} and \citeA{lai2008srh}.

\subsubsection{Training bathymetry generation}
There are different ways to generate the bathymetry for training cases. Among them, the most popular way uses a stochastic process which can embed some prior knowledge on the bed. In this work, the training bathymetry is randomly drawn from a 2D Gaussian process. To control the properties of the generated bathymetry, such as overall bed elevation range and feature size, the parameters of the Gaussian process can be tuned. Additionally, if ground-truth bed elevations at certain locations are available, they can be assimilated into the Gaussian process prior to the draw. Following \citeA{Landon2014}, this work uses the Radial-Basis Function (RBF) as the 2D Gaussian kernel, which has the form of 
\begin{equation}
  k(\mathbf{x}, \mathbf{x'}) = \sigma_{z_b}^2 \exp \left[ - \left( \frac{\Delta x^2}{2l_x^2} + \frac{\Delta y^2}{2l_y^2} \right) \right]
\end{equation}
where $\mathbf{x}$ and $\mathbf{x'}$ are two arbitrary points in the domain; $\Delta x$ and $\Delta y$ are their distance in $x$ and $y$ directions, respectively; $\sigma_{z_b}$ is the prior variance; $l_x$ and $l_y$ are the length scales in $x$ and $y$ directions, respectively.

\begin{figure}[htp]
\centering
    \includegraphics[width=1.0\textwidth]{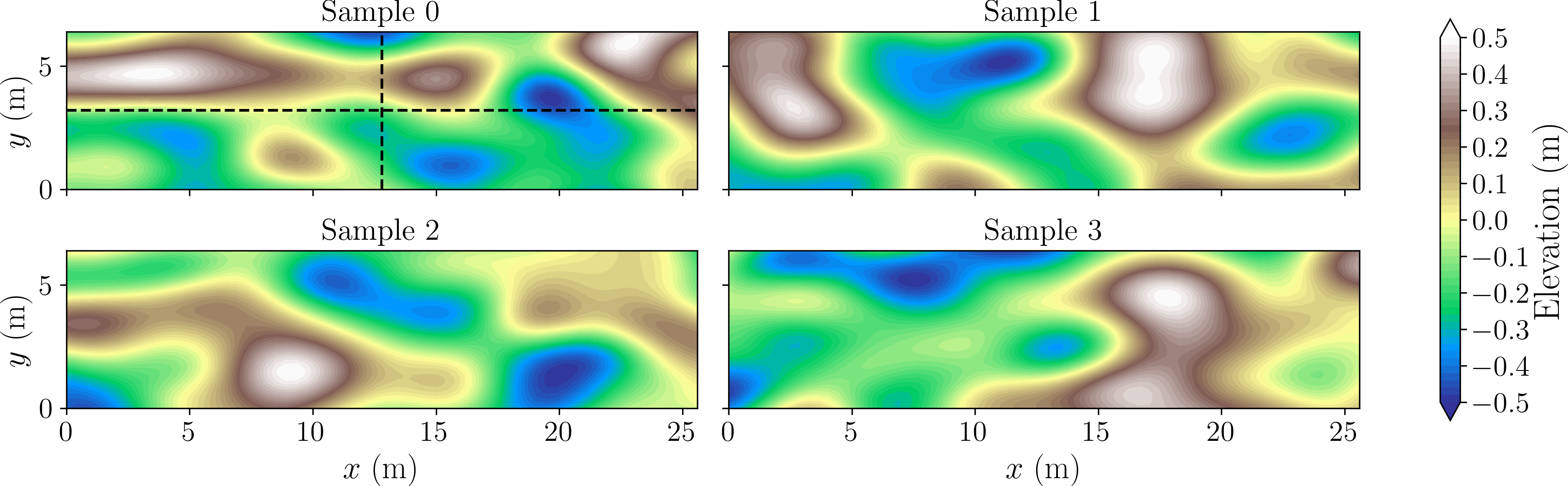}
    \caption{Four example synthetic bathymetries generated with 2D Gaussian process. The two dashed lines on Sample 0 are the two profiles, longitudinal and cross-sectional, used later in this paper. Flow is from left to right.}\label{Fig:example_bathymetry}
\end{figure}

The parameters for the kernel should be set based on some basic prior information about the bathymetry. For demonstration purpose, this work generated hypothetical bathymetries as training data. The bathymetries were firstly generated on a unit square. The prior variance $\sigma_{z_b}$ has a value of 0.5. The length scales $l_x$ and $l_y$ have the values of 0.1 and 0.2, respective. Then, the bathymetries on the unit square were mapped to a rectangular domain with a length of 25.6 m and a width of 6.4 m. These parameters were set such that the generated bed elevation is in the range of -0.5 m to 0.5 m and there is approximately one or two significant bed features (bedform-like) in the simulation domain. Three thousand bathymetires were generated in this work. Figure~\ref{Fig:example_bathymetry} shows four samples and flow is from left to right. For example, Sample 0 shows one elongated bar on the top and one hump at the outlet. There is also a deep pool near the outlet hump. Sample 0 will be referenced later as an example for some detailed analysis.

\subsubsection{Flow data generation, parameter tuning, and implementation}
Three thousand simulation cases were run with the generated bathymetries, out of which 80\%, 15\%, and 5\% were used for training, validation and testing, respectively. All cases were run with the same conditions except the bathymetry. Steady flow in the rectangular channel was simulated. The inflow discharge was 3 m$^3$/s and the downstream was set with a fixed water surface elevation of 1 m. The computational domain was discretized into 1580 triangles with an average size of 0.5 m. A constant Manning's roughness $n$ of 0.03 was used. All simulations were run until steady state. All simulation results and input bathymetries were then sampled on a grid of 32 rows and 128 columns, which is proportional to the width-to-length ratio of the channel. The grid resolution is double that of the computational mesh to ensure the capture of bathymetry and flow field details. 

The performance of surrogate model and the inversion highly depends on the choice of hyperparameters. The optimal values for these hyperparameters were obtained through manual tuning, which is manageable because of the relative simplicity of the hypothetical cases in this work. Automatic and more intelligent hyperparameter optimization can be performed if the cases are more complex. The CNN-based surrogate model used in this work contains two convolution layers in the encoder and four transpose convolution layers in the decoder for each of the outputs. The fully connected layer for the code has a length of 1,024. Other parameters of the surrogate model can be found in Table~\ref{Tab:CNN_details}.

\begin{table}[htp]
\caption{Details of the CNN-based surrogate model}
\footnotesize
\begin{tabular}{lp{8cm}p{2.5cm}}
\hline
 Unit & Layers & Size ($H$, $W$, $N_f$) or number of neurons \\
\hline
 Input $z_b$ &  & (32,128,1) \\
 Encoder & Conv2D, 128 filters of size (8,8), stride (8,8), padding=``same''  & (4,16,128)  \\
 Encoder & Conv2D, 512 filters of size (4,4), stride (4,4), padding=``same''  & (1,4,512)  \\
 Encoder & Flatten   & (2048,1,1)  \\
 Fully connected & Dense   & (1024)  \\
 Reshape & Reshape   & (1,1,1024)  \\
 Decoder & Conv2DTranspose, 512 filters of size (8,8), stride (8,8), no padding  & (8,8,512)  \\
 Decoder & Conv2DTranspose, 256 filters of size (8,2), stride (8,2), no padding  & (64,16,256)  \\
  Decoder & Conv2DTranspose, 1 filter of size (2,2), stride (2,2), no padding  & (128,32,1)  \\
  Decoder & Permute with order (2,1,3) & (32,128,1)  \\
\hline
\end{tabular}\label{Tab:CNN_details}
\end{table}

During the surrogate training process, to update the weights and biases in the neural network, an initial learning rate was set at $10^{-3}$. A learning rate scheduler was utilized to reduce the learning rate by a factor of 0.5 when the validation loss did not decrease over 5 epochs. One hundred epochs with a mini-batch size of 10 were performed until both training and validation losses were simultaneously minimized. For inversion, a fixed learning rate of 0.01 was used to perform 1,000 steps of iterations which has been proven to be sufficient to obtain converged solutions.

The surrogate model and the inversion method were implemented in Tensorflow \cite{tensorflow2015-whitepaper} with Keras API \cite{chollet2015keras}. The training of the surrogate model and the inversion were performed on Amazon AWS with a NVIDIA K80 GPU card. The simulations with the physics-based model SRH-2D were performed on a desktop with an Intel Core 3.40GHz CPU. The 3,000 SRH-2D simulations took about 24 hours to finish. The training of the surrogate model took about 2 hours. Each inversion took about 1 minute. The code and data generation scripts used in this work can be accessed at \url{https://github.com/psu-efd/dl4HM/tree/main/examples/bathymetry_inversion_2D}. Because of the large number of simulations needed, and to automate the data generation and processing, the python package, Python-based Hydraulic Modeling Tools, $pyHMT2D$, was used to control SRH-2D modeling runs and transform the results to the inputs and outputs of the neural network \cite{pyHMT2D}. Source code of $pyHMT2D$ can be found on GitHub (\url{https://github.com/psu-efd/pyHMT2D}). 

\section{Results and discussions}\label{sect:result}

\subsection{Performance of surrogate model}\label{sect:surrogate_performance}
The premise of accurate inversion using surrogate model is the accuracy of the surrogate model itself. In this work, the surrogate model was evaluated with three metrics, i.e., the MSE defined in Eqn.~\ref{eqn:L_prediction}, the root-mean-square error (RMSE) of absolute error and relative error. The relative error for a grid point is defined as the ratio of the absolute error to the maximum of the predicted and ground-truth values (to avoid division by zero problem). Taking $u$ as an example, the RMSE of absolute error ${e}_{{m,u}}$ and relative error ${e}_{{r,u}}$ for each case are defined as follows:
\begin{equation}\label{eqn:rmse_absolute_error}
{e}_{{m,u}} = \sqrt{ \frac{1}{N_{points}}  \sum_{i=1}^{N_{points}}  \left(\hat{{u}}-{u} \right)^2 }
\end{equation} 
\begin{equation}\label{eqn:average_relative_error}
{e}_{{r,u} }=  \frac{1}{N_{points}}  \sum_{i=1}^{N_{points}}  \frac{\left|\hat{{u}}-{u}\right|}{\max(\left|\hat{u}\right|,\left|u\right|)} 
\end{equation} 
where $N_{points}$ is the number of points in each test case. Then the mean, max, and standard deviation of the two errors for all test cases are calculated.

\begin{figure}[htp]
\centering
    \includegraphics[width=0.6\textwidth]{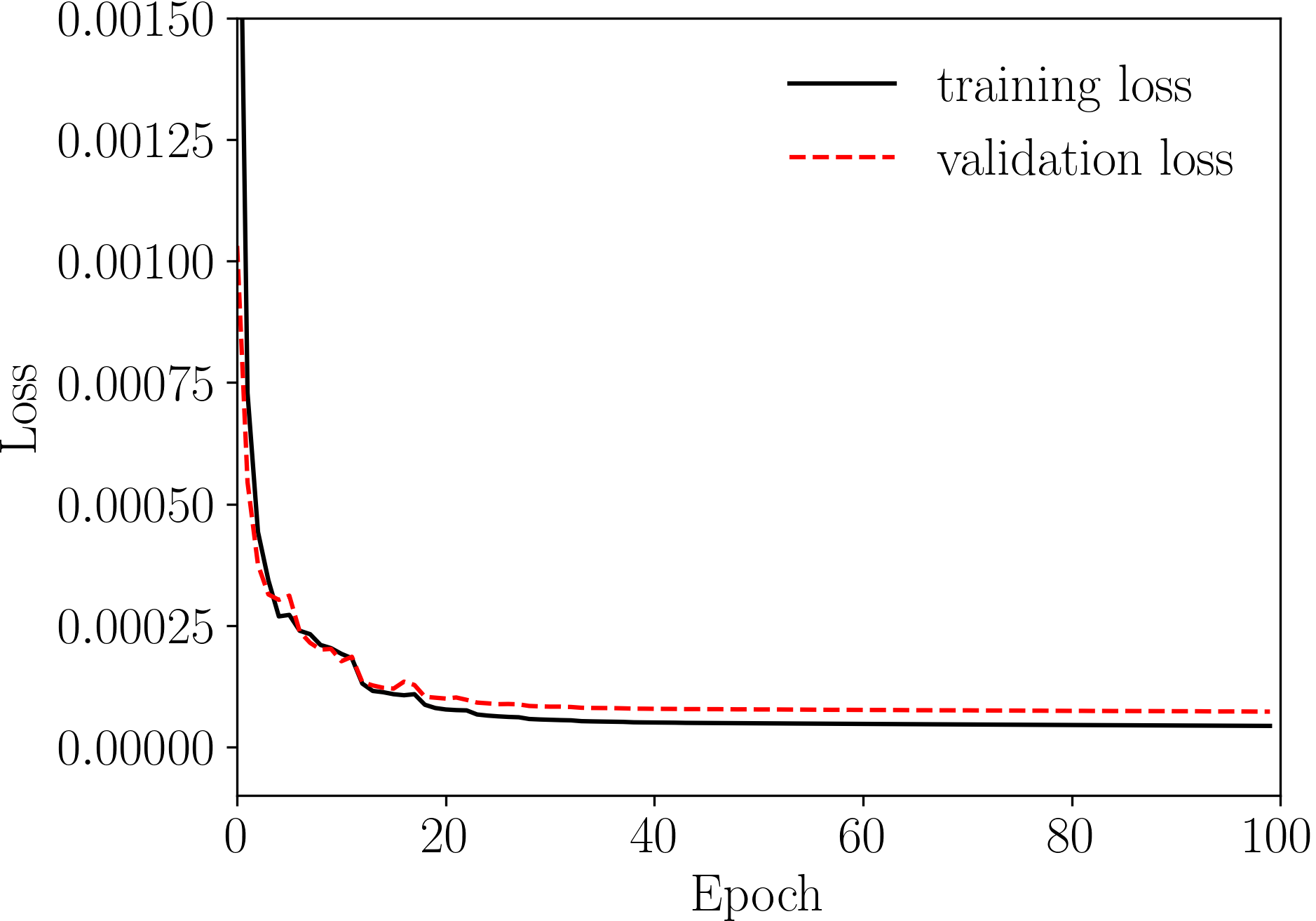}
    \caption{Training and validation losses.}\label{Fig:training_validation_losses}
\end{figure}

Figure~\ref{Fig:training_validation_losses} shows the history of the training and validation MSE losses over 100 epochs. Both losses were driven to their minimums and there was no significant over-fitting because the validation loss is close to the training loss. This shows the architecture of the CNN surrogate and its hyperparameters are proper for the given dataset.

\begin{figure}[htp]
\centering
    \includegraphics[width=1.1\textwidth]{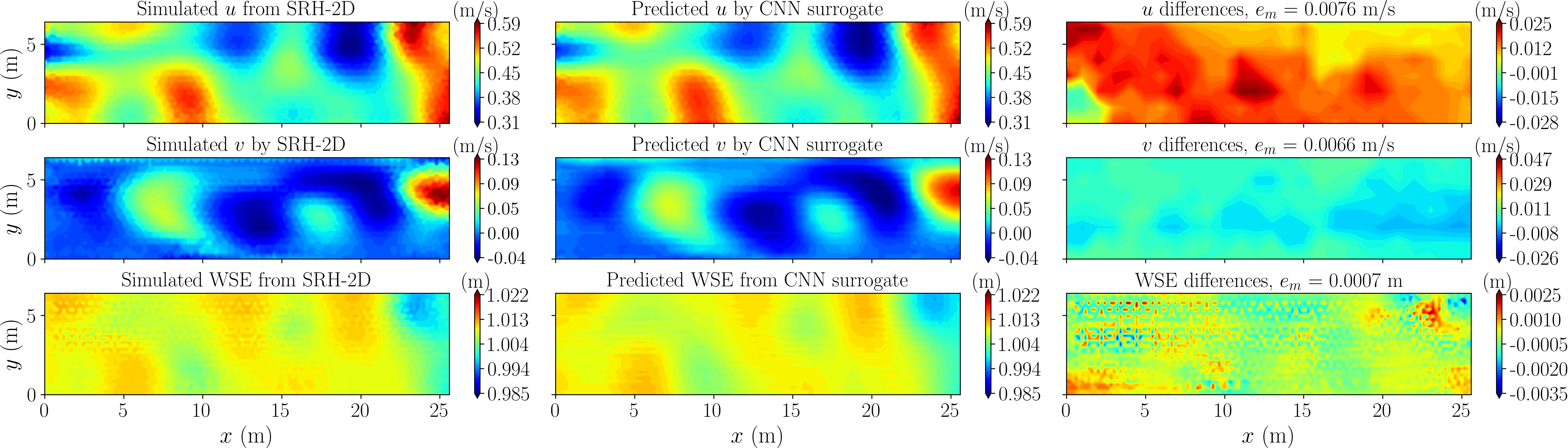}
    \caption{An example case showing the performance of CNN surrogate model. The bathymetry for this case is the ``Sample 0'' shown in Fig.~\ref{Fig:example_bathymetry}. The first and second columns show the contours of velocity components $u$, $v$, and $WSE$ from SRH-2D and CNN surrogate, respectively. The third column shows the differences.}\label{Fig:surrogate_prediction}
\end{figure}

The surrogate model proposed in this work is of high accuracy and properly captures the input-output dynamics embedded in the SWEs solver. Figure~\ref{Fig:surrogate_prediction} shows an example case comparing the results from SRH-2D and the surrogate. The bathymetry of this case is the ``Sample 0'' shown in Fig.~\ref{Fig:example_bathymetry}. The result shows that the difference between the results of PBM and surrogate is small. For this particular case, the RMSEs of absolute error for $u$, $v$ and $WSE$ are 0.0076 m/s, 0.0066 m/s, and 0.0007 m, respectively. 

The prediction error for $WSE$ is smaller, in both absolute and relative senses, than the errors for the two velocity components. This is not surprising because the $WSE$ field is typically much smoother and has less variations than the velocity field. In fact, because of the hydrostatic assumption in SWEs, the water depth $h$, which directly controls $WSE$, is equivalent to the pressure in the Navier-Stokes equations. In the Navier-Stokes equations, the pressure is governed by an elliptic Poisson's equation, which has self-smoothing effect over time and space for proper initial and boundary conditions. 

The relatively small prediction error in $WSE$ may have some implications for inversion. For a given bathymetry, it is easy for the surrogate model to make more accurate $WSE$ predictions than the velocity. In other words,  the total error is dominated by the velocity errors. If this is true, then during inversion the $WSE$ error does not contribute as much as the velocity error. Therefore, it is possible to drop $WSE$ and only use velocity components $u$ and $v$ for the inversion. This is good news because that means we need less information to invert the bathymetry. In practice, it is difficult to obtain velocity and $WSE$ at the same time. Although the CNN surrogate model has three output branches ($u$, $v$ and $WSE$), we do not need to use all of them to do the inversion. More discussion will be presented in the inversion result section.   

The spatial distributions of prediction errors shown in Figure~\ref{Fig:surrogate_prediction} do not show any clear trend in where large errors are located. For example, locations of large prediction errors in the velocity components do not seem to be coincident with either low or high of bathymetry and velocity themselves. These observations apply to all test cases. It might be beneficial in future research to investigate what drives the spatial distribution of prediction error from the CNN and whether it has implications for inversion.

The statistics of the flow prediction errors for all 150 test cases are shown in Table~\ref{Tab:surrogate_accuracy}. The statistics include the mean, max, and standard deviation (std). The prediction errors for all test cases are small, except the relative error for the velocity component $v$, which has a mean value of about 39.6\% and a maximum of about 61.6\%. This high relative error should not cause any alarm because it is due to the small magnitude of $v$ in the flow results. The absolute error for $v$ is reassuringly very small.   
 
\begin{table}[htp]
\caption{RMSE of absolute error and relative error to show the accuracy of surrogate model for all 150 test cases.}
\small
\centering 
\begin{tabular}{c c c c c c c} 
\hline
\multirow{2}{*}{Flow variable} &\multicolumn{3}{c}{\parbox{4cm}{\centering \linespread{1}\selectfont RMSE of absolute error \\ (m/s for $u$ and $v$ and m for $WSE$)}}&\multicolumn{3}{c}{Relative error (\%)}\\ %\cline{2-7} 
\cmidrule(lr){2-4} \cmidrule(lr){5-7}
{} & mean  & max & std & mean & max & std \\
\hline
$u$ & 0.0079  &  0.0169  &  2.27e-03  &  1.3155  &  2.7425  &  0.3786 \\
\hline
$v$ & 0.0066  &  0.0246  &  2.06e-03  &  39.6035  &  61.5791  &  4.7272 \\
\hline
WSE & 0.0007  &  0.0035  &  2.98e-04  &  0.0554  &  0.2060  &  0.0204 \\
\hline
\end{tabular}
\label{Tab:surrogate_accuracy}
\end{table}

\begin{figure}[htp]
\centering
    \includegraphics[width=0.8\textwidth]{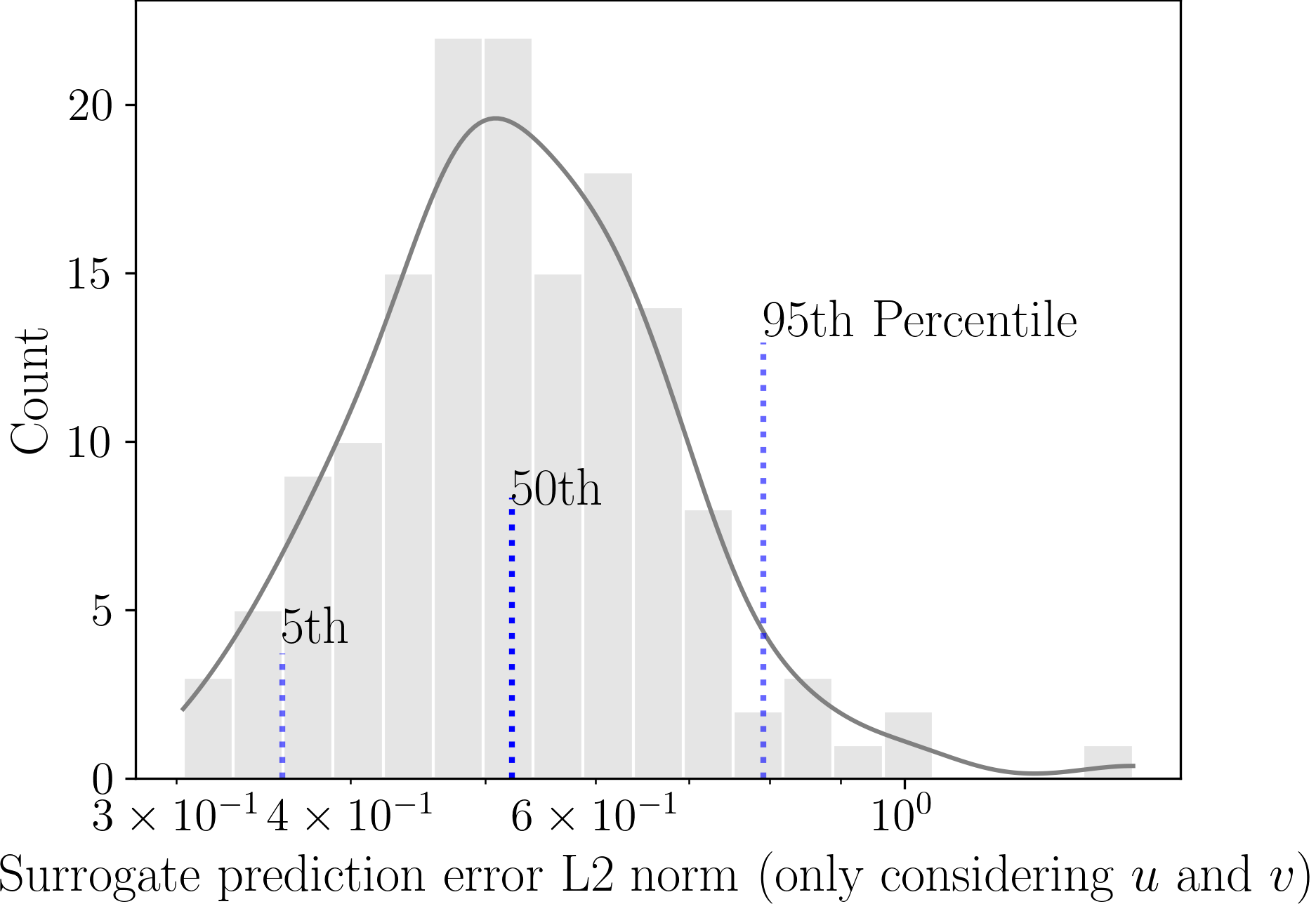}
    \caption{Histogram of the L2 norms of surrogate prediction errors for all 150 test cases. Here the L2 norm only has the contributions from $u$ and $v$ prediction errors.}\label{Fig:surrogate_prediction_l2norm_histogram}
\end{figure}

The accuracy of the surrogate model was further analyzed with the distribution of the L2 norm of the prediction error $\mathbf{\epsilon}$ in Eqn.~\ref{eqn:surrogate}. The histogram, the 5th, 50th, and 95th percentiles of the L2 norms for all 150 test cases are shown in Fig.~\ref{Fig:surrogate_prediction_l2norm_histogram}. The L2 norm of surrogate error $\mathbf{\epsilon}$ is used in the determination of the inversion loss parameters $\alpha_{value}$ and $\alpha_{slope}$. Here, the L2 norm only has the contributions from $u$ and $v$ prediction errors because only velocity is typically used for inversion. From the figure, it is clear that the surrogate model error is in a range (approximately from 3$\times$10$^{-1}$ to 1.0) and each case's error is different. Its implication for inversion is discussed next.

\subsection{Performance of bathymetry inversion}

\subsubsection{Inversion parameter determination}

The performance of inversion also heavily depends on the parameters. The values of relevant parameters for inversion are listed in Table~\ref{Tab:inversion_hyperparameters}. Indeed, some of the parameters have physical meanings and should reflect the specific problem setup. For example, the mean and amplitude of bed elevation value and slope should be set with our prior knowledge of the bathymetry. In this case, the normalized bed elevation should be in the ranges of [-0.5, 0.5]. The bed slopes in $x$ and $y$ directions should be in [-0.08,0.08] and [-0.15, 0.15], respectively. In practice, this prior information has to be obtained from other means such as survey or historical maps.

\begin{table}[htp]
\caption{Hyperparameters for inversion}
\centering 
\begin{tabular}{lc}
\hline
Hyperparameter & Value \\
\hline
Learning rate $\alpha_{inversion}$ & 0.01 \\
\parbox{8cm}{\linespread{1}\selectfont Value loss regularization factor $\alpha_{value}$} & case dependent \\
\parbox{8cm}{\linespread{1}\selectfont Bed elevation mean for value loss regularization $x_{c,value}$} & 0.0 \\
\parbox{8cm}{\linespread{1}\selectfont Bed elevation amplitude for value loss regularization $a_{value}$} & 0.5 \\
\parbox{8cm}{\linespread{1}\selectfont Slope loss regularization factor $\alpha_{slope}$} & case dependent \\
\parbox{8cm}{\linespread{1}\selectfont Slope mean for slope loss regularization in both $x$ and $y$ directions $x_{c,slope}$} & 0.0 \\
\parbox{8cm}{\linespread{1}\selectfont Slope amplitude for slope loss regularization in $x$ direction $a_{slopex}$} & 0.08 \\
\parbox{8cm}{\linespread{1}\selectfont Slope amplitude for slope loss regularization in $y$ direction $a_{slopey}$} & 0.15 \\
\hline
\end{tabular}\label{Tab:inversion_hyperparameters}
\end{table}

Among all the parameters, $\alpha_{value}$ and $\alpha_{slope}$ defined in Eqn.~\ref{eqn:inversion_L_total} are the most difficult to be determined. Their values effectively balance the surrogate prediction error and regularizations in the total inversion loss $L_{total}$. Among other things, they depend on the surrogate model error $\mathbf{\epsilon}$. For inverse problems, the ``L-curve'' criterion is often used to determine their values \cite{Aster2013}. When plotted on a log-log scale, the curve of model prediction loss vs. regularization loss for a linear problem often has a ``L'' shape, hence the name. However, for nonlinear problems, the curve often shows more complicated shapes. In addition, the ``L-curve'' criterion only works for one parameter. In this work, we found that the parameter of bed elevation value, $\alpha_{value}$, is not sensitive. Its function is to tamp down bed elevations exceeding the specified range. In addition, the smoothing effect of the slope regularization also has certain effect to reduce the extreme bed elevations. Through trial and error, a value of 0.1 for $\alpha_{value}$ was found to be appropriate. Therefore, we will focus on $\alpha_{slope}$ next. 

\begin{figure}[htp]
\centering
    \includegraphics[width=0.8\textwidth]{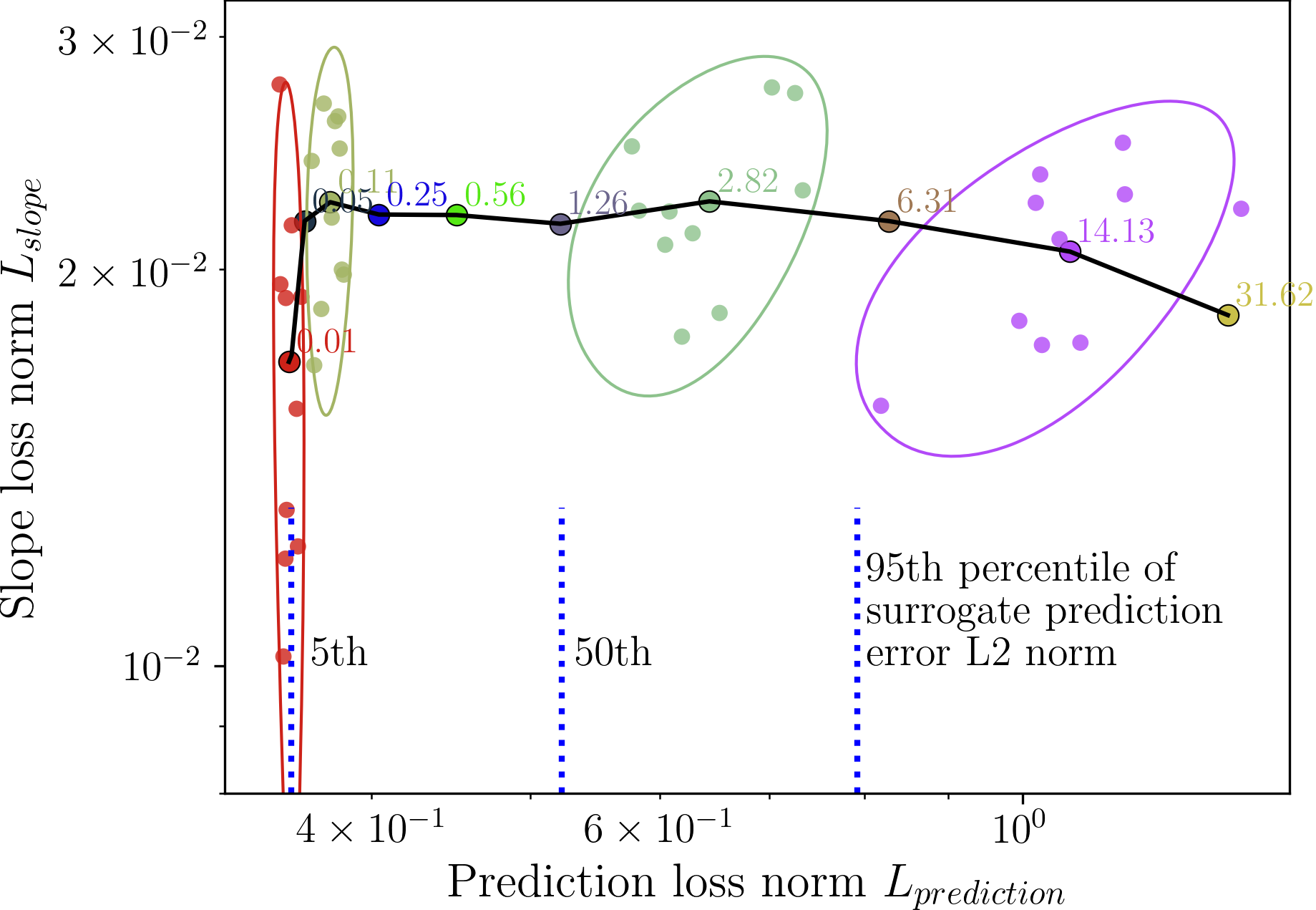}
    \caption{Determination of slope loss parameter $\alpha_{slope}$ for an example case. This case corresponds to ``Sample 0'' in Fig.~\ref{Fig:example_bathymetry}. The scatter markers are colored by their corresponding $\alpha_{slope}$ values, which are annotated in the figure. For each $\alpha_{slope}$ value, 11 inversions were performed with different initial guesses of the bed. The mean of all 11 inversions for each $\alpha_{slope}$ value are plotted as a larger marker with edge. All the mean points are connected to show the ``L-curve''. For the 11 cases of each $\alpha_{slope}$ value, the confidence ellipse with two standard deviations is also drawn to show the clustering. The percentiles of the surrogate prediction errors are also shown.}\label{Fig:L_curve}
\end{figure}

The value of $\alpha_{slope}$ is determined through a heuristic approach inspired by the ``L-curve'' criterion for linear problems. For each case, we performed the inversion using 10 different values of $\alpha_{slope}$ in the range of [0.01, 31.6]. For each $\alpha_{slope}$ value, 11 starting beds $\hat{z}_b^{0}$ were randomly generated. So for each case, we performed 110 inversions. Their corresponding flow prediction losses $L_{predition}$ and slope regularization losses $L_{slope}$ are plotted in Fig.~\ref{Fig:L_curve}. To reduce clutter, only a subset of the data is plotted. The mean of all 11 inversions for each $\alpha_{slope}$ value is plotted as a larger marker with edge. All the mean points are connected to show the ``L-curve''. The figure shows that the curve does not have an ``L'' shape due to nonlinearity. For the 11 cases of each $\alpha_{slope}$ value, the confidence ellipse with a radius of two standard deviations is also drawn to show the clustering.

Unlike in linear cases where there is a clear corner point on the ``L-curve'' for both $L_{predition}$ and $L_{slope}$ to reach their respective minimums, there is no such point for the nonlinear case shown in Fig.~\ref{Fig:L_curve}. Here, we propose a heuristic approach which is based on the following concurrent requirements:
\begin{itemize}
  \item minimize prediction loss $L_{predition}$. The figure shows that the prediction loss decreases as $\alpha_{slope}$ decreases. This is expected because in the limit of $\alpha_{slope}$=0, the surrogate model makes the most accurate prediction within the bound of its accuracy. Based on this requirement, we may choose a small $\alpha_{slope}$ value such as 0.01.
  \item minimize the scatter (uncertainty) in prediction loss $L_{predition}$. The figure shows that the scatter in prediction loss $L_{predition}$, measured by the horizontal size of the confidence ellipse, decreases as $\alpha_{slope}$ decreases. This is again as expected because the surrogate model performs best with no extra terms added to the loss function. Based on this requirement, we may again choose a small $\alpha_{slope}$ value such as 0.01.
  \item minimize the scatter (uncertainty) in slope loss $L_{slope}$. The figure shows that the scatter in slope loss $L_{slope}$, measured by the vertical size of the confidence ellipse, is the largest when $\alpha_{slope}$ has a value of 0.01. For other values, the scatter is comparable.   
\end{itemize}

Balancing all the above requirements, the $\alpha_{slope}$ value of 0.1 seems most reasonable. In addition, the prediction losses $L_{predition}$ corresponding to this $\alpha_{slope}$ value are within the 5th and 95th percentiles of the surrogate model prediction accuracy. Results beyond the range bounded by the 5th and 95th percentiles do not make sense. They either overfit or underfit the $\alpha_{slope}$ value to drive the surrogate model out of its accuracy range.

\subsubsection{Inversion result evaluation}
This section shows the bathymetry inversion results. Like in many previous researches, the inversion discussed in this section only used the velocity $u$ and $v$ as the input. From the practical point of view, this is reasonable because velocity data can be obtained relatively easily. The effects of inclusion or exclusion of $WSE$ in the inversion is discussed later. 

\begin{figure}[htp]
\centering
    \includegraphics[width=1.2\textwidth]{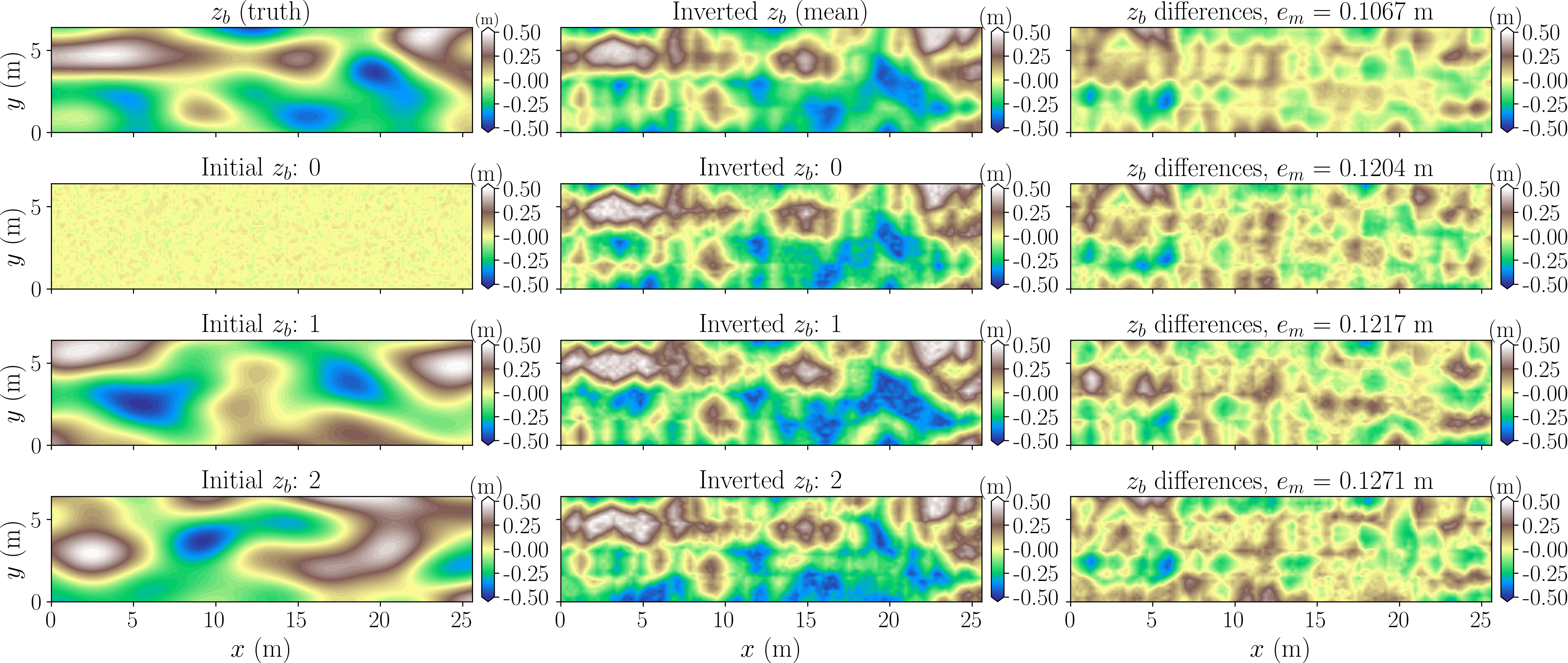}
    \caption{Example inversion results for $z_b$. The top row, from left to right, shows the $z_b$ truth, mean of all inverted $z_b$ fields from all initial conditions, and the differences between truth and inverted beds. The rest rows show three individual examples of inversion results started from different initial beds. Note that for the second row the initial bed is flat with random white noise.}\label{Fig:zb_inversion}
\end{figure}

As an example, Fig.~\ref{Fig:zb_inversion} shows the inversion results for one of the cases. All these inversion cases are in the test dataset, not in the training and validation datasets. Note here the truth bathymetry is again the ``Sample 0'' shown in Fig.~\ref{Fig:example_bathymetry}. The top row of Fig.~\ref{Fig:zb_inversion}, from left to right, shows the $z_b$ truth, mean of all inverted $z_b$ fields from all eleven initial bathymetries, and the differences between truth and mean inverted beds. The rest rows of Fig.~\ref{Fig:example_bathymetry} show three individual examples of the inversion from three different initial bathymetries, including the one from the initial flat bed with random white noise. The results show that all inversions for this particular case converged to similar, though subtly different, final bathmetires. 

The inverted bathymetries not only qualitatively, but also quantitatively, recover the truth. For the example shown in Fig.~\ref{Fig:zb_inversion}, all inversions in the eleven ensemble capture the elongated bar on the top and the hump at the outlet. In addition, they also capture the deep pool feature near the outlet. Quantitatively, the RMSE of the inversion error is about 0.08 m for the mean and about 0.09 m for individual inversions. The slight improvement by the mean shows the benefits of using an ensemble, instead of a single inversion, because of the uncertainties.

\begin{figure}[htp]
\centering
    \includegraphics[width=0.6\textwidth]{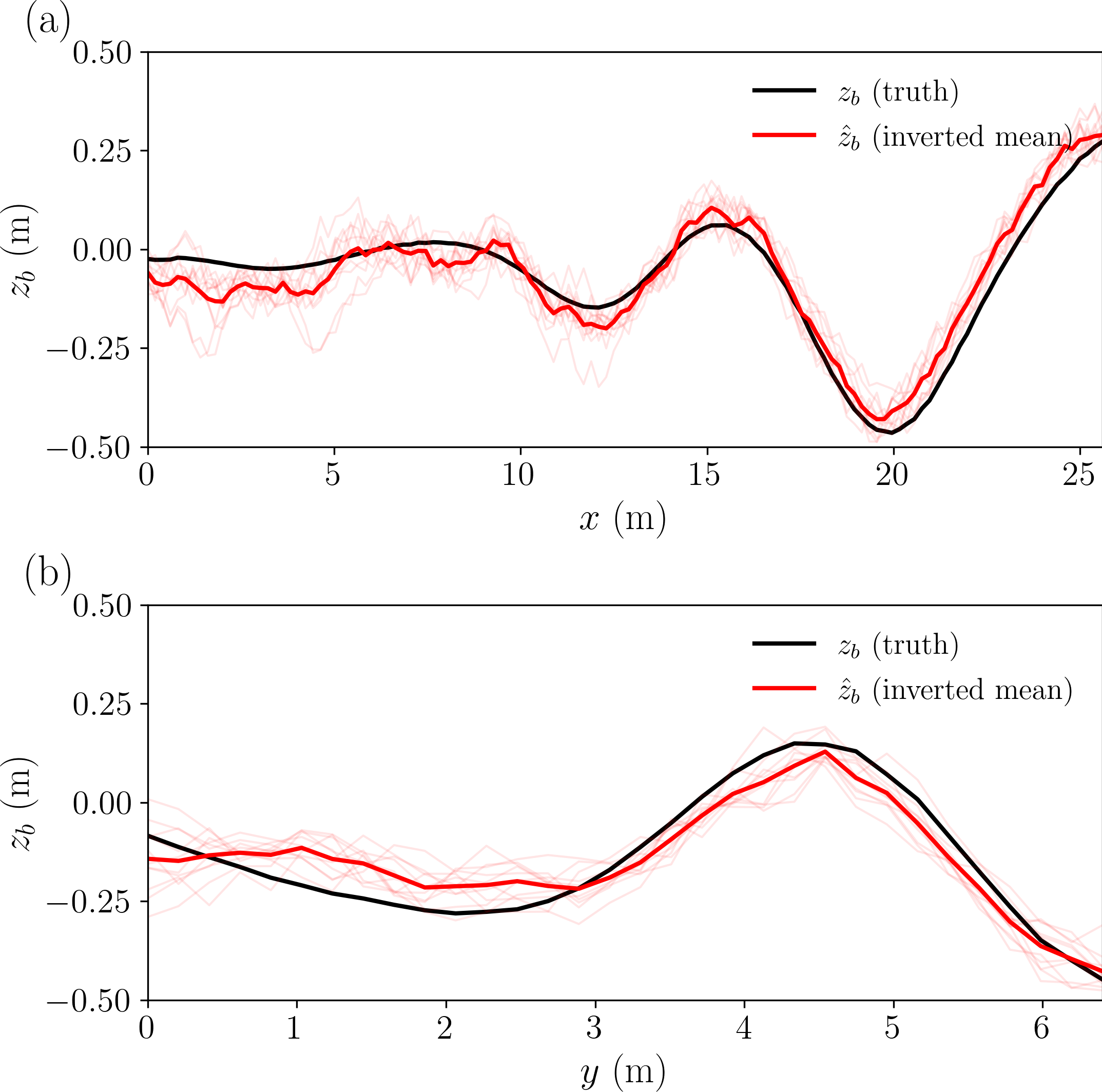}
    \caption{Profiles of the inverted bathymetry for the case shown in Fig.~\ref{Fig:zb_inversion}: (a) Longitudinal profile in the middle of the channel, (b) Cross-sectional profile at half channel length. The light profiles are the results inverted from different initial beds.}\label{Fig:zb_inversion_profiles}
\end{figure}

To further appreciate the efficacy and uncertainty of the inversion method proposed in this work, Fig.~\ref{Fig:zb_inversion_profiles} shows the profiles of the inverted beds for the same case shown in Fig.~\ref{Fig:zb_inversion}. The locations of the two profiles, one longitudinal and one cross-sectional, are shown as the black dashed lines on bathymetry ``Sample 0'' in Figure~\ref{Fig:example_bathymetry}. The two profiles highlighted with heavy weight lines are for the truth and the mean of all inversions. The ensemble of inverted bed profiles is blended in the background with light weight lines to show the band of uncertainties. The mean of inverted bed profiles generally follows the profiles of truth. The uncertainty band approximately bounds the truth for both longitudinal and cross-sectional profiles, which is an evidence for the reliability of the inversion result. 

\subsubsection{Inversion process analysis}
\begin{figure}[htp]
\centering
    \includegraphics[width=\textwidth]{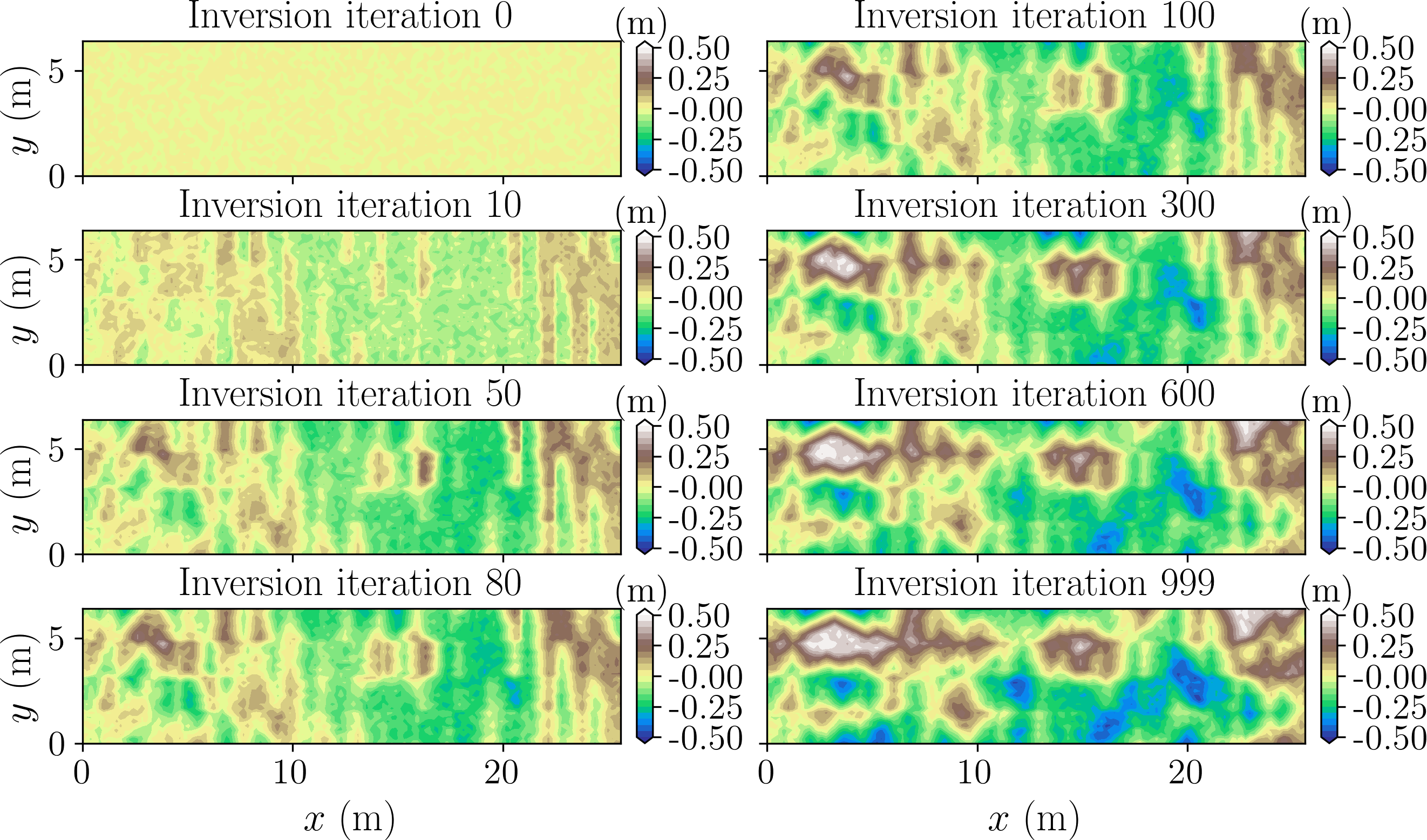}
    \caption{Example inversion process for the case shown in Fig.~\ref{Fig:zb_inversion}. This inversion used the initial flat bed with random noises.}\label{Fig:inversion_process}
\end{figure}

\begin{figure}[htp]
\centering
    \includegraphics[width=0.6\textwidth]{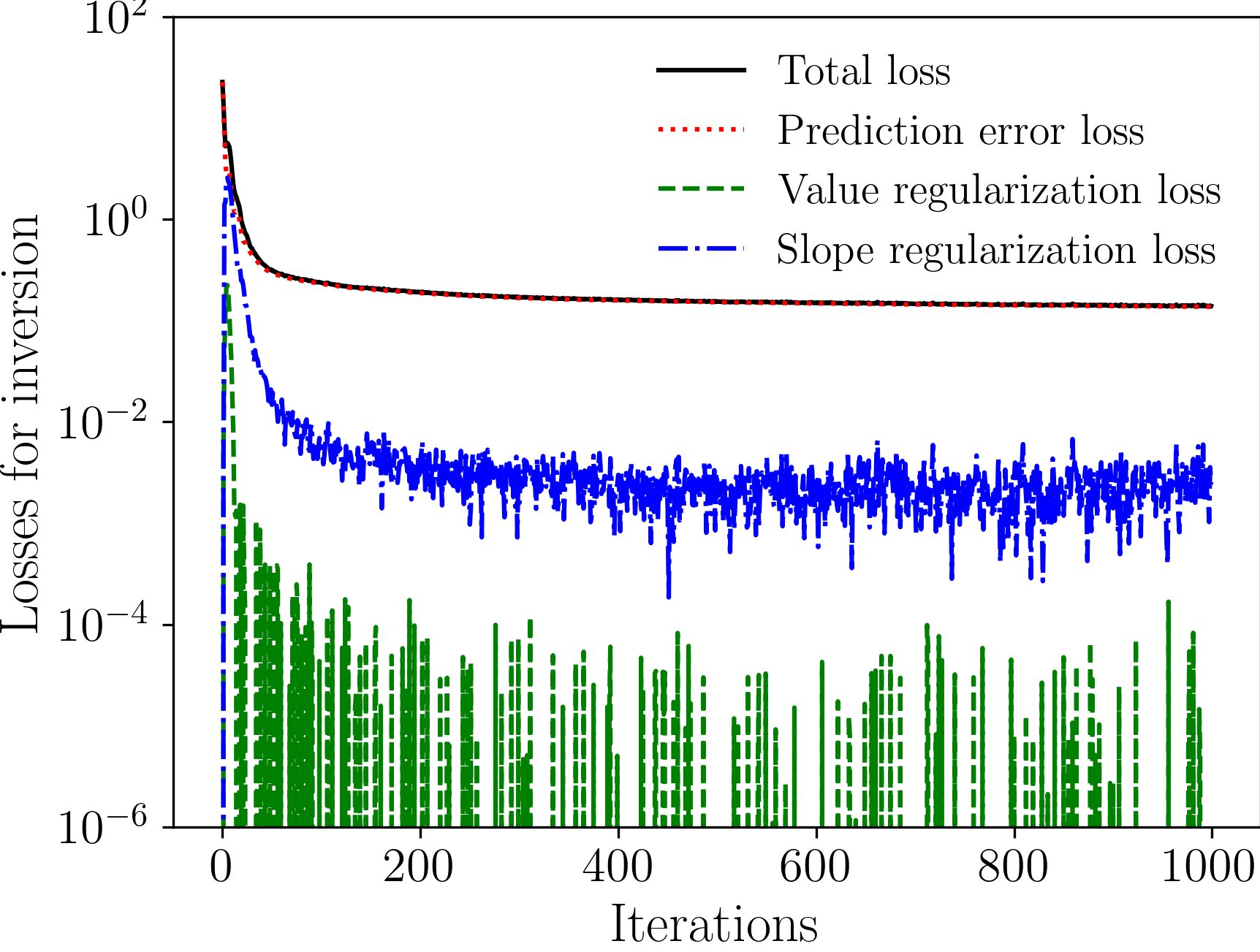}
    \caption{Histories of inversion total loss and its three components as functions of iteration steps. The prediction error loss is due to the error in the predicted flow field. The other two losses are due to bed elevation value and slope regularizations. The case is the same as that in Fig.~\ref{Fig:inversion_process}.}\label{Fig:inversion_loss_hisotry}
\end{figure}

Some detailed analysis was performed on the inversion process to shed light on how the iterative inversion algorithm finds the solution. Figure~\ref{Fig:inversion_process} shows the inverted bathymetry at eight iteration steps for the same example case shown in Fig.~\ref{Fig:zb_inversion}. This inversion example started using the initial flat bed with random noises. An animation of this inversion example is in the Supplementary Information. To aid the analysis, the total inversion loss and its components are plotted in Fig.~\ref{Fig:inversion_loss_hisotry}.

The inversion process clearly shows two stages. The first stage is between Iteration 0 to about 100. This initial stage is characterized by the rapid change of bed with blocks or bands such that it can quickly evolve to a state that can describe the overall landscape of the target bathymetry. This can be easily observed in the left column of Fig.~\ref{Fig:inversion_process} where the inverted beds at Iteration 0, 10, 50, and 100 are shown. The initial inverted beds show some hint on the two bed bars at the top and the outlet, as well as the deep pool at the outlet (the truth bahtymetry is ``Sample 0'' in Fig.~\ref{Fig:example_bathymetry}). The first stage can also be easily identified in the loss history plotted in Fig.~\ref{Fig:inversion_loss_hisotry}. During this stage, the total loss and its three components all drops quickly. It is noted that at Iteration 0, the regularization losses due to bed elevation value and slope are zero. However, after the first iteration which has not taken into account the regularizations yet, the inverted bed has significant violations of the imposed bounds for bed elevation and slope. Thus, the losses for value and slope regularizations have a sudden jump at Iteration 1. The inverted beds from the first several iterations are far from the truth. Thus, the loss due to errors in the predicted flow field is also high. The continuation of the inversion iterations drives down all the losses efficiently because the total loss is very responsive to the change of bed, i.e., the gradient $\partial L_{total}(z_b)/\partial z_b$ has very large magnitude in the first stage.

\begin{figure}[htp]
\centering
    \includegraphics[width=\textwidth]{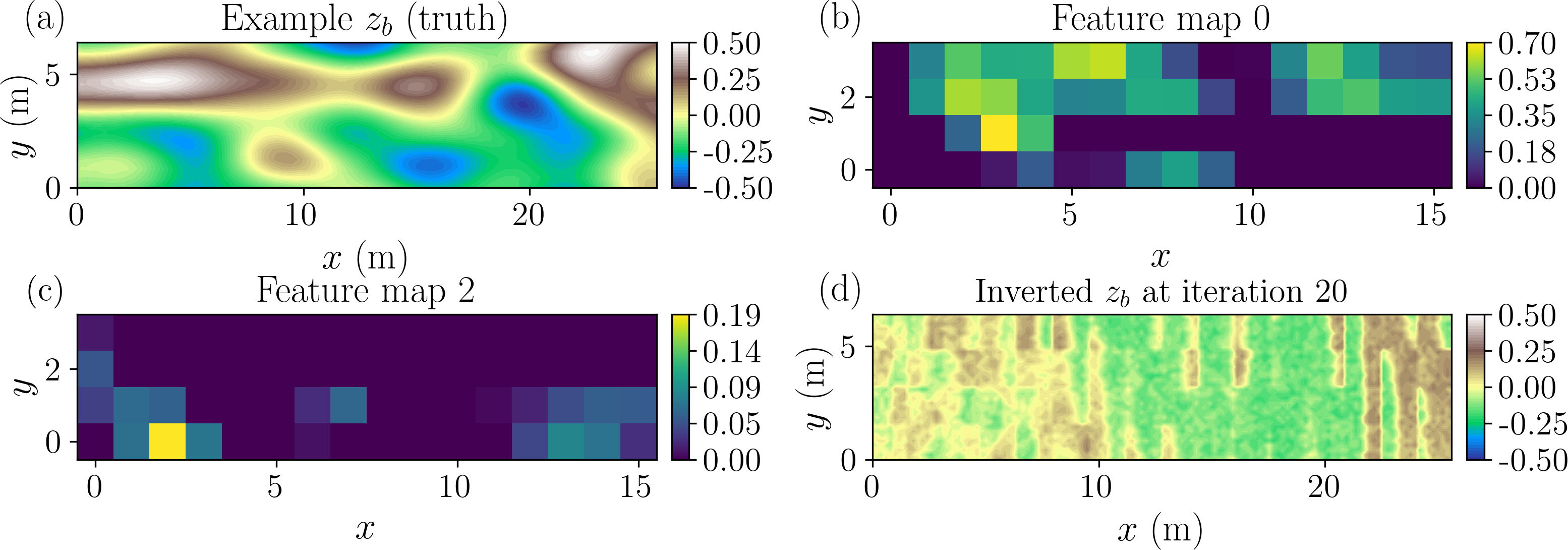}
    \caption{Feature maps and inversion process: (a) Truth bathymetry $z_b$, (b) One example feature map in the first convolution layer corresponding to the truth bathymetry, (c) Another example feature, and (d) Inverted the bathymetry at one early iteration showing similar pattern of feature map 0.}\label{Fig:feature_maps_inversion}
\end{figure}

The blocks or bands of the inverted beds at the beginning of the inversion process is due to the convolutional nature of filters in the encoder. Specifically, the length scale of the blockiness or band is directly proportional to the filter and stride sizes in the convolutional layer, especially the first one. Some portion of information on the bathymetry, in particular the spatial scales smaller than the scale defined by the filter and stride sizes, is lost during encoding process. This is one inherent limitation of using image-based regressions such as CNN. To demonstrate, Fig.~\ref{Fig:feature_maps_inversion} shows one example bed, two feature maps in the first convolution layer of the encoder corresponding to this bed, and the inverted bed at one early iteration. The shown feature maps are two representatives out of the 128 feature maps in the first convolution layer. One can observe that the two feature maps roughly capture some bed peaks and depressions. It is also clear that the inverted bed at the shown iteration has  similar pattern of feature map 0. Intuitively, the inversion is initially guided by the large-scale, main features of the bed embedded in the feature maps.

One could argue that to reduce the blockiness or bands of the initial inverted beds, we can reduce the sizes of filter and stride, and add more feature maps and layers in the CNN surrogate model. However, this naive approach will also increase the complexity of the CNN architecture and the trainable parameters will increase exponentially. For a given training dataset, the surrogate model with increased complexity will quickly overfit and make the surrogate model prediction useless. Therefore, there is a trade-off between the accuracies of forward prediction and inversion using CNN-based surrogate models. It is beyond the scope of this work to investigate whether we can and if so, how to overcome this dilemma. One potential solution is to use surrogate models for point-to-point, instead of image-to-image, predictions. In other words, we can build surrogate models without the use of CNN technique. Recently, such surrogate model has been proposed in \citeA{song2021surrogate}.

Fortunately, the blockiness or band of inverted beds at the beginning of inversion is not a big concern. The banded bed will be smoothed out in the second stage which happens roughly after Iteration 100. As shown in the right column in Fig.~\ref{Fig:inversion_process}, the blocks and edges of the inverted bed are gradually smoothed out. In this process, more small-scale bathymetric details emerge to better depict the final inverted bed. All these are driven by the continued minimization of the total loss, which needs to minimize all three loss components. From Fig.~\ref{Fig:inversion_loss_hisotry}, it is important to note that in this second stage (after Iteration 100), the prediction loss $L_{prediction}$ stays almost constant. The value regularization loss oscillates with negligibly small values, which indicates that inverted bed elevation is well within the imposed bound. What makes significant changes is the slope regularization loss. Indeed, the bed slope changes mostly happen at the edge of the large-scale blocks resulted from the first stage. The slope regularization resembles the detailing and finishing of a rough sculpture.

\subsection{Effects of inversion regularization}
Regularization is the key to achieving usable inversion results. This section will show the effects of the two regularizations proposed in Section~\ref{sect:inversion_process_method}. Four cases will be compared, i.e., with both value and slope regularizations, with slope regularization only, with value regularization only, and with no regularization. 

\begin{figure}[htp]
\centering
    \includegraphics[width=1.2\textwidth]{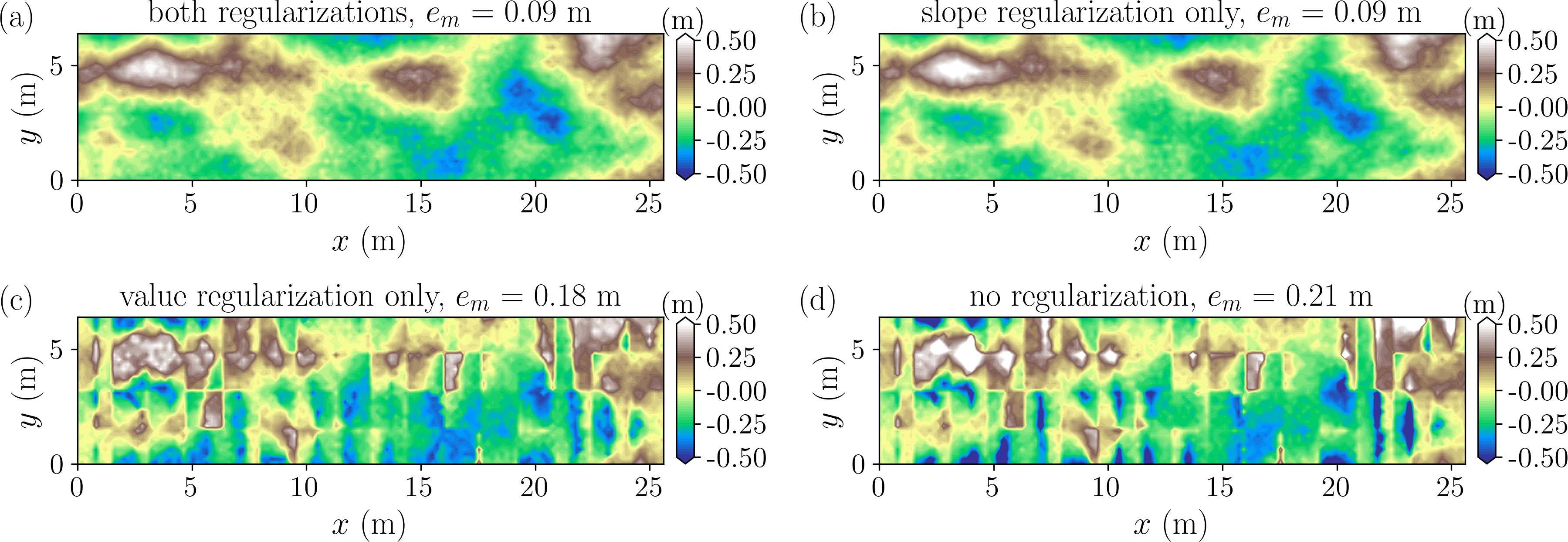}
    \caption{Effects of inversion loss regularization on inverted bathymetries: (a) with both slope and value regularizations, (b) with only slope regularization, (c) with only value regularization, and (d) with no regularization. The RMSE of inverted $z_b$ is also reported.}\label{Fig:zb_inversion_contours_regularization_effects}
\end{figure}

\begin{figure}[htp]
\centering
    \includegraphics[width=\textwidth]{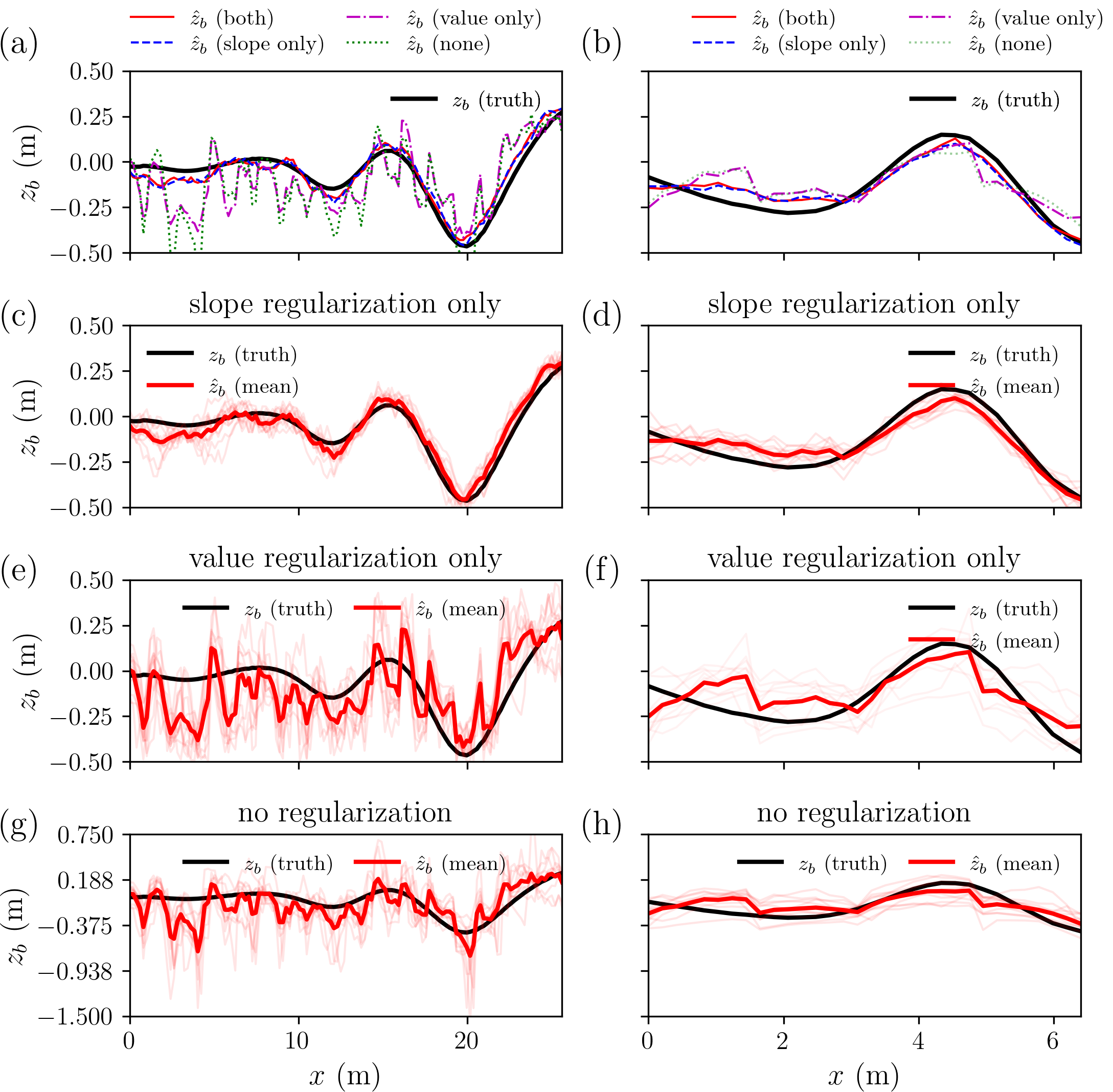}
    \caption{Effects of inversion regularization on inverted bed profiles. The first column is for the longitudinal profiles and the second for cross-sectional profiles. (a) and (b) show the mean profiles from all regularization scenarios in comparison with the truth. (c) and (d) show the profiles inverted with slope regularization only. (e) and (f) show the profiles inverted with value regularization only. (g) and (h) show the profiles inverted with no regularization.}\label{Fig:zb_inversion_profiles_regularization_effects}
\end{figure}

Figure~\ref{Fig:zb_inversion_contours_regularization_effects} shows the contours of inverted bathymetries for the four cases. The RMSE of inverted $z_b$ is also reported on the figure. Figure~\ref{Fig:zb_inversion_profiles_regularization_effects} plots the longitudinal and cross-sectional profiles of the inverted bathymetries. Again, the truth bathymetry is the ``Sample 0'' in Fig.~\ref{Fig:example_bathymetry}. The results show that without proper regularization, the inversion produces suboptimal or even garbage solutions in comparison with the truth. With slope regularization, the inverted bathymetry is close to the truth. However, because of no constraint on value, the inverted bed elevation is beyond the specified range in many places (see the portion of bright white and deep blue pixels beyond the range of colorbar in Fig.~\ref{Fig:zb_inversion_contours_regularization_effects}(b)). With only value regularization, the inverted bathymetry shows more blockiness and noisiness. The noisiness can also be observed in the profiles shown in Fig.~\ref{Fig:zb_inversion_profiles_regularization_effects}. The worst result among the four is the one with no regularization, whose RSME is the highest. 

Although not plotted, the results also show the prediction losses $L_{prediction}$ for both cases of slope regularization only and value regularization only are comparable. That means the inverted bathymetries shown in Fig.~\ref{Fig:zb_inversion_contours_regularization_effects}(b) and (c) are both admissible solutions if the flow prediction loss is the only metric. This is a clear evidence of the non-uniqueness for the bathymetry inversion problem. To shrink the solution space and nudge the inversion toward a usable solution, physical constraints in the form of inversion loss regularizations must be utilized. The result show that the constraint on the bed elevation value itself using the zeroth-order Tikhonov regularization is not sufficient. Additional constraint on the bed slope, i.e., the smoothness of the inverted bathymetry, is also critical. For practical applications, these additional constraints on value and slope should be adjusted based on prior knowledge or belief of the bathymetry to be inverted.

\subsection{Inversion uncertainty}
Because of the non-uniqueness and ill-posed nature of bathymetry inversion, uncertainty quantification is of great interest for practical purpose. To demonstrate the use of the inversion method proposed in this work, a simple uncertainty study was performed. The inversion input, i.e., velocity, was augmented with 10\% uncertainty by adding random perturbations. Two hundred inversions were performed with the randomly perturbed velocity fields and then statistics were calculated for the inverted bathymetries.

\begin{figure}[htp]
\centering
    \includegraphics[width=0.6\textwidth]{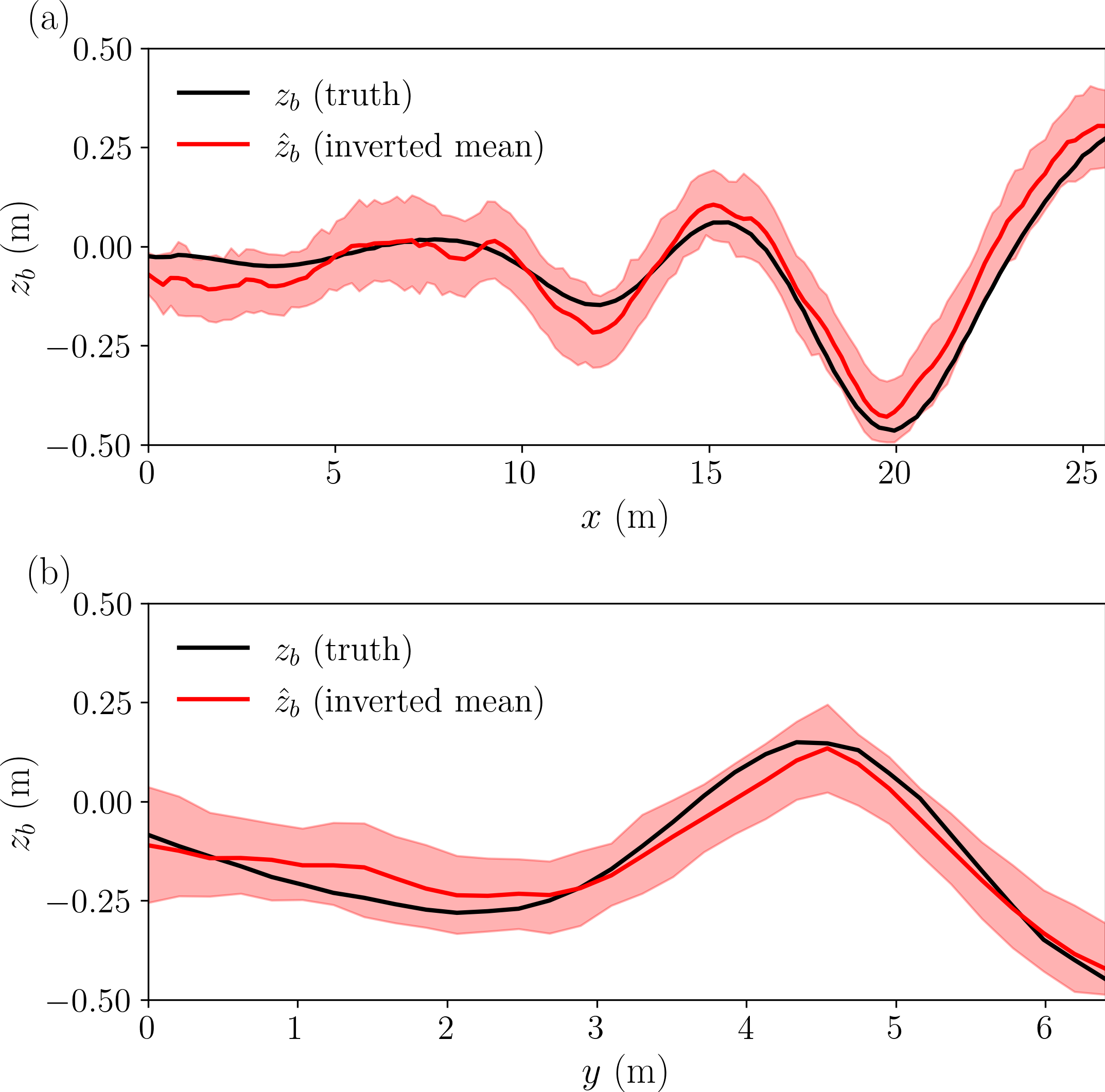}
    \caption{Profiles of the inverted bed with 10\% uncertainty added to the velocity field: (a) Longitudinal profile in the middle of the channel, (b) Cross-sectional profile at half channel length. The 95\% confidence intervals around the mean are shown as the shaded areas.}\label{Fig:zb_inversion_profiles_uv_uncertainty}
\end{figure}

Figure~\ref{Fig:zb_inversion_profiles_uv_uncertainty} shows the mean inverted longitudinal and cross-sectional profiles with uncertain velocity fields. The 95\% confidence intervals around the mean profiles are shown as shaded areas. The truth bed profiles are shown as black solid lines. The case uses the ``Sample 0'' bathymetry in Fig.~\ref{Fig:example_bathymetry} as the truth. The result shows that the truth bed profiles are bounded by the 95\% confidence intervals, indicating the efficacy of the proposed inversion method and the appropriateness of the hyperparameter values. 

\subsection{Effects of CNN surrogate architecture}
The architecture of CNN surrogate model affects the forward prediction. It is also important to know how the architecture affects the inversion. This section discusses two aspects of this effect. 

One aspect is that so far we only used two outputs of the surrogate model, i.e., two velocity components $u$ and $v$, in the inversion. The question is then whether it has any advantage to design a CNN surrogate which only has two outputs for $u$ and $v$. To distinguish, we denote the surrogate with three outputs as NN$_{(u,v,WSE)}$ and the one with only two outputs as NN$_{(u,v)}$. Theoretically, the training of the two surrogate models with the same dataset will result in different parameters (weights and biases) in the neural nets. Consequently, the gradient of $\partial L_{total}(z_b)/\partial z_b$ in Eq.~\ref{eqn:zb_inversion} will be different. The surrogate model NN$_{(u,v)}$ was trained using the same dataset except that it only has two outputs for $u$ and $v$, not $WSE$ (see Fig.~\ref{Fig:CNN_structure}). 

The second aspect is whether the inclusion of $WSE$ in the inversion, in conjunction with $u$ and $v$, makes any difference. In Section~\ref{sect:surrogate_performance}, we already discussed the relatively small prediction error for $WSE$ in comparison with those for $u$ and $v$. We hypothesize that the inclusion of $WSE$ in inversion does not have significant contribution for the inversion, at least for the problem defined by the dataset in this work.  

\begin{figure}[htp]
\centering
    \includegraphics[width=\textwidth]{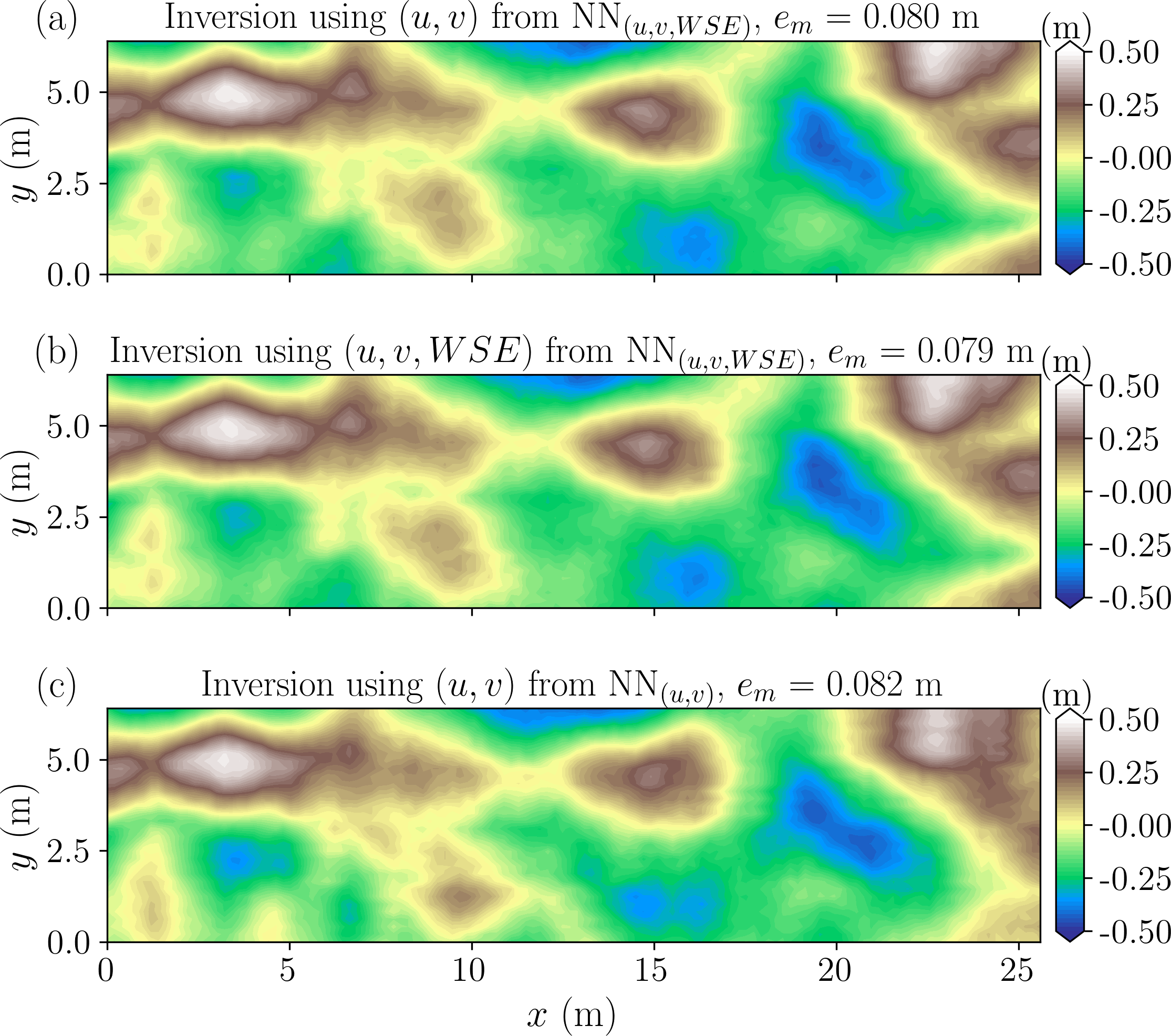}
    \caption{Effects on inversion using different CNN surrogate architectures and different inversion input: (a) inversion using ($u$, $v$) from NN$_{(u,v,WSE)}$, (b) inversion using ($u$, $v$, $WSE$) from NN$_{(u,v,WSE)}$, and (c) inversion using ($u$, $v$) from NN$_{(u,v)}$.}\label{Fig:zb_inversion_contours_loss_comp_cnn_structure}
\end{figure}

Based on above discussion, we compare three different approaches, namely inversions using ($u$, $v$) from NN$_{(u,v,WSE)}$, ($u$, $v$, $WSE$) from NN$_{(u,v,WSE)}$, and ($u$, $v$) from NN$_{(u,v)}$. Figure~\ref{Fig:zb_inversion_contours_loss_comp_cnn_structure} shows the inverted bathymetries using the three approaches. Each subplot shows the mean of the bathymetries inverted from eleven different starting guesses of the bed. Although slightly different, all three approaches produce similar bathymetries with comparable RMSE. For the surrogate model NN$_{(u,v,WSE)}$, the inversions with and without $WSE$ produce almost identical results, which confirms our hypothesis regarding the importance of $WSE$ for this problem. Whether this conclusion can be generalized is unclear at this point. It may so happen that $u$ and $v$ are sufficient to do the inversion and $WSE$ is redundant for this particular problem. For a different problem where $WSE$ can contribute more new information, its inclusion may be necessary. 

Judging by the comparable RMSE of inverted bathymetry, the inversion using ($u$, $v$) from NN$_{(u,v)}$ has no advantage than that using ($u$, $v$) from NN$_{(u,v,WSE)}$. Future work needs to investigate the generalizability of this conclusion. Again, for cases where $WSE$ can contribute unique information, the inclusion of $WSE$ might be necessary and therefore inversion using ($u$, $v$) from NN$_{(u,v)}$ might perform poorly.

\section{Conclusion}
Using a CNN-based surrogate model for a shallow water equations solver, a bathymetry inversion method is developed based on the gradient conveniently calculated with neural network's automatic differentiation. The surrogate model uses a shared-encoder and separate-decoder architecture, which can successfully capture the dynamics between input (bathymetry) and output (flow field). To do the inversion, new regularizations have to be used for both the bed elevation value and bed slope. The new regularizations embed the prior knowledge or belief on the bathymetry to be inverted. Without these regularizations, especially the slope regularization, the inverted bathymetry shows substantial blockiness and noisiness. 

One of the difficulties in nonlinear inversion problem is the lack of clearly defined approach for determining the parameters of regularization losses. In this work, we found the $\alpha_{slope}$ parameter for slope loss is more important than $\alpha_{value}$ for bed elevation value. Using the ``L-curve'' criterion, a heuristic approach was proposed with some guiding requirements. This approach produces optimal $\alpha_{slope}$ values and good inversion results. 

The inversion process has two distinctive stages. The first is characterized by rapid changes of bed in blocks and bands such that it can quickly evolve to a state that can represent the overall landscape of the target bathymetry. The inversion loss due to flow prediction error quickly converges to its minimum during this stage. The blockiness of the inverted intermediate beds correspond to the sizes of filters and strides for the feature maps of convolution layers in encoder. During the second stage, the regularizations for bed elevation value and slope gradually smooth out the blocks and bands of bed and add more details to the inverted bathymetry. In this second stage, the inversion loss due to flow prediction error stays almost constant. It is the local adjustment of bed slope and elevation that drive the bathymetry toward its final solution. In other words, it is the two regularizations selecting the most probable solution, out of many, to fit the imposed physical constrains on the bathymetry. 

The investigation on the CNN surrogate architecture reveals that different options on the CNN surrogate (NN$_{(u,v,WSE)}$ vs. NN$_{(u,v)}$) and whether to include $WSE$ in the inversion yield comparable bathymetries. The fact that the use of NN$_{(u,v,WSE)}$ and NN$_{(u,v)}$ produces similar results may be due to shared-encoder and separate decoder architecture of the surrogate. In this way, the separate decoders for $u$, $v$ and $WSE$ have minimum interference. For NN$_{(u,v,WSE)}$, the inclusion of $WSE$, in conjunction with $u$ and $v$, is redundant, at least for the problem solved in this work. The root cause of this is that the prediction error for $WSE$ is much smaller than those for velocity, and thus it contributes less in the total inversion loss. This conclusion may not be true in other cases where the prediction error for $WSE$ responds strongly to the variation of bathymetry.   

The use of surrogate model greatly reduces the computational cost of forward modeling runs during the inversion. However, this saving is at the cost of offline surrogate training time. In future work, this cost may be reduced using other approaches. For example, instead of surrogate models, an alternative option is to implement the physics-based forward models using machine learning platform and languages such as PyTorch and Tensorflow. In this way, all the operations on the inputs are recorded and automatic differentiation can be carried out directly using backpropagation. However, this alternative currently faces at least two challenges. One is the upfront cost of implementing forward models using a new language. The second is the complexity and computational cost of automatic differentiation even if we can implement physics-based models using these platforms. Solvers of SWEs using traditional methods such as the finite volume method involve tremendous amount of calculations which have to be recorded internally for automatic differentiation purpose. 

\section{Open Research}
A permanent copy of code, data, and scripts used for this work has also been achieved in the CUAHSI's HydroShare:
\url{http://www.hydroshare.org/resource/357c3c413622460a91d29cc61d0ba084}. The latest version of code and data generation scripts can also be accessed at \url{https://github.com/psu-efd/dl4HM/tree/main/examples/bathymetry_inversion_2D}. 

\acknowledgments
This work is supported by a seed grant from the Institute of Computational and Data Sciences at the Pennsylvania State University.

\bibliography{references}

\end{document}